\documentclass[paper]{JHEP}
\usepackage{epsfig}
\def\beq{\begin{equation}}
\def\beqn{\begin{eqnarray}}
\def\eeq{\end{equation}}
\def\eeqn{\end{eqnarray}}
\def\abs#1{\left|#1\right|}
\def\bentarrow{\:\raisebox{1.3ex}{\rlap{$\vert$}}\!\longrightarrow}
\def\colket#1{|#1\rangle}
\def\colbra#1{\langle#1|}
\def\bom#1{{\mbox{\boldmath $#1$}}} 
\newcommand\f[2]{\frac{#1}{#2}} 
\newcommand\sss{\scriptscriptstyle}
\newcommand\ep{\epsilon}
\newcommand\vep{\epsilon}
\newcommand\stepf{\Theta}
\newcommand\epem{e^+e^-}
\newcommand\as{\alpha_{\sss S}}
\newcommand\asu{\alpha_{\sss S}^u}
\newcommand\gs{g_{\sss S}}
\newcommand\E{{\cal E}}
\newcommand\ENLO{{\cal E}_{\sss NLO}}
\newcommand\ENNLO{{\cal E}_{\sss NNLO}}
\newcommand\vp{\vec{p}}

\newcommand\toplab{{\cal T}}
\newcommand\To{{\bf 1}}
\newcommand\Topo{{\bf 1}\!\otimes\! {\bf 1}}
\newcommand\Tt{{\bf 2}}
\newcommand\Am{{\cal M}_m}
\newcommand\An{{\cal M}_n}
\newcommand\Anone{{\cal M}_{n+1}}
\newcommand\Antwo{{\cal M}_{n+2}}
\newcommand\Tm{{\cal M}_m^{(0)}}
\newcommand\Tn{{\cal M}_n^{(0)}}
\newcommand\TnStar{{\cal M}_n^{(0)^\star}}
\newcommand\Tnone{{\cal M}_{n+1}^{(0)}}
\newcommand\TnoneStar{{\cal M}_{n+1}^{(0)^\star}}
\newcommand\Tntwo{{\cal M}_{n+2}^{(0)}}
\newcommand\LOm{{\cal M}_{m}^{(1)}}
\newcommand\LOn{{\cal M}_{n}^{(1)}}
\newcommand\LOnStar{{\cal M}_{n}^{(1)^\star}}
\newcommand\LOnone{{\cal M}_{n+1}^{(1)}}
\newcommand\LOnoneStar{{\cal M}_{n+1}^{(1)^\star}}
\newcommand\LTm{{\cal M}_{m}^{(2)}}
\newcommand\LTn{{\cal M}_{n}^{(2)}}
\newcommand\LTnStar{{\cal M}_{n}^{(2)^\star}}
\newcommand\cm{{\cal M}}
\newcommand\cmt{{\cal M}^{(0)}}
\newcommand\cmo{{\cal M}^{(1)}}
\newcommand\cmd{{\cal M}^{(2)}}
\newcommand\pt{p_{\sss T}}
\newcommand\qt{q_{\sss T}}
\newcommand\kt{k_{\sss T}}
\newcommand\kti{k_{{\sss T}i}}
\newcommand\kto{k_{{\sss T}1}}
\newcommand\ktd{k_{{\sss T}2}}
\newcommand\ktt{k_{{\sss T}3}}
\newcommand\ktmu{k_{{\sss T}\mu}}
\newcommand\ktnu{k_{{\sss T}\nu}}
\newcommand\tfmo{\tilde{f}_{-1}}
\newcommand\tfmt{\tilde{f}_{-2}}
\newcommand\trrmo{\widetilde{rr}_{\!-1}}
\newcommand\trrmt{\widetilde{rr}_{\!-2}}
\newcommand\mydot{\!\cdot\!}
\newcommand\CA{C_{\sss A}}
\newcommand\CF{C_{\sss F}}
\newcommand\TR{T_{\sss R}}
\newcommand\nf{n_{\sss F}}
\newcommand\bq{\bar{q}}
\newcommand\qp{q^\prime}
\newcommand\bqp{\bar{q}^\prime}
\newcommand\qqq{q^{}}
\newcommand\rrtbrc{\qp_1\bqp_2\qqq_3}
\newcommand\rrobrc{\qp_1\bqp_2}
\newcommand\psp{\Phi}
\newcommand\pspb{\overline{\Phi}}
\newcommand\Mx{{\rm max}}
\newcommand\bK{{\bom K}}
\newcommand\bKtw{{\bom K}^{(2)}}
\newcommand\bKth{{\bom K}^{(3)}}
\newcommand\bVtw{{\cal V}^{(2)}}
\newcommand\bT{{\bom T}}
\newcommand\Ph{{\hat P}}
\newcommand\CpC{C\oplus C}
\newcommand\Euler{\gamma_{\sss E}}
\def\lra{\leftrightarrow}
\newcommand\nn{\nonumber}
\newcommand\la{\langle}
\newcommand\ra{\rangle}
\newcommand\ymax{y_{\rm max}}
\newcommand\Li{{\rm Li}}

\preprint{GEF-TH-13/2004}
\title{Subtraction at NNLO}
\author{Stefano Frixione\\
  INFN, Sezione di Genova,
  Via Dodecaneso 33, I-16146 Genoa, Italy\\
  E-mail: \email{Stefano.Frixione@cern.ch}}
\author{Massimiliano Grazzini\\
  Dept. of Physics, Theory Division, CERN, CH-1211 Geneva 23, Switzerland and\\
  INFN, Sezione di Firenze,
  Via G.~Sansone 1, I-50019 Sesto F., Florence, Italy\\
  E-mail: \email{Massimiliano.Grazzini@cern.ch}}
\abstract{
We propose a framework for the implementation of a
subtraction formalism at NNLO in QCD, based on an observable- and
process-independent cancellation of infrared singularities.
As a first simple application, we present the calculation of the
contribution to the $\epem$ dijet cross section 
proportional to $\CF\TR$.}

\begin{document}
\section{Introduction\label{sec:intro}}
The hard dynamics of processes involving hadrons is nowadays remarkably
well described by QCD predictions. An important role has been played
in this achievement by the ability to compute the relevant reactions 
to next-to-leading order (NLO) accuracy, a mandatory step in view 
of the large value of the coupling constant $\as$. Although fairly
successful in their phenomenological applications, NLO predictions 
are affected by uncertainties that can be of the order of a few
tens of percent, typically estimated by varying the renormalization and 
factorization scales. This is a consequence of the fact that, in some cases,
NLO corrections are numerically as important as the
leading order (LO) contributions. By increasing the accuracy of
the perturbative predictions, through the computation of the
next-to-next-to-leading order (NNLO) contributions, one would
certainly reduce the size of the uncertainties, and obtain a firmer
estimate of the rates~\cite{Giele:2002hx}.
Such a task, however, implies finding the
solution of a few highly non trivial technical problems. Recently, 
two major bottlenecks have been cleared: two-loop functions with up to 
four legs have been computed (with zero or one massive leg)~\cite{twoloop},
and so have the three-loop Altarelli-Parisi kernels~\cite{Moch:2004pa},
which opens the way to {\em exact} NNLO PDF fits.

The present situation is thus fairly similar to that of the early 80's.
Back then, the computation of the
two-loop Altarelli-Parisi kernels~\cite{Curci:1980uw}, 
and the ability to compute all the tree-level and one-loop functions
for specific processes, left open the problem of achieving explicitly
the cancellation of soft and collinear singularities, as predicted
by the Kinoshita-Lee-Nauenberg (KLN) theorem for inclusive,
infrared-safe observables. Early approaches pioneered
the subtraction~\cite{Ellis:1980wv}
and phase-space slicing~\cite{Fabricius:1981sx} techniques,
for computing analytically the divergent 
part of the real corrections, in the context of the prediction for a
given observable; a different observable required a novel computation.
Later, it was realized that the KLN cancellation could be 
proven in an observable- and process-independent manner, which allows
one to predict any observable (for which the relevant matrix elements 
can be computed) to NLO accuracy without actually using the observable
definition in the intermediate steps of the 
computation~\cite{Giele:1991vf}--\cite{Nagy:1996bz}.
Apart from being very flexible, these universal methods have the virtue of
clarifying the fundamental structure of the soft and collinear
regimes of QCD.

Observable- and process-specific calculations have certain advantages over 
universal formalisms; the possibility of exploiting the observable definition
and the kinematics of a given reaction in the intermediate steps of the 
computation generally leads to more compact expressions to integrate. 
This is evident if we consider the fact that the first NNLO results for 
total rates~\cite{sigmatot,Hamberg:1990np} pre-date the universal NLO 
formalisms by some years. More recently, the first calculation of a 
rapidity distribution at NNLO has been performed using these techniques,
for single vector boson
hadroproduction~\cite{Anastasiou:2003yy,Anastasiou:2003ds}.

In the last two years a fair amount of work has been carried out in 
the context of process-specific calculations. The structure of NNLO 
infrared-singular contributions necessary to implement the subtraction 
method has been discussed in the case of 
$e^+e^-\to 2~{\rm jets}$~\cite{Gehrmann-DeRidder:2004tv},
and the $\CF^3$ contribution to $e^+e^-\to 3~{\rm jets}$
has been computed~\cite{Gehrmann-DeRidder:2004xe}.
Moreover, an observable-independent method~\cite{Anastasiou:2003gr},
based on sector decomposition~\cite{sector}, has been recently proposed. 
This approach allows one to handle and cancel infrared singular
contributions appearing in the intermediate steps of an NNLO calculation
in a fully automatic (numerical) way, and thus significantly differs from the 
semi-analytical approaches of refs.~\cite{Giele:1991vf}--\cite{Nagy:1996bz}.
This method has been applied to the computations of 
$e^+e^-\to 2~{\rm jets}$~\cite{Anastasiou:2004qd} and Higgs 
hadroproduction~\cite{Anastasiou:2004xq} cross sections.

One may argue that, since the number of two-loop amplitudes computed so 
far is limited, process-specific (but observable-independent) computations
such as those of 
refs.~\cite{Gehrmann-DeRidder:2004xe,Anastasiou:2004qd,Anastasiou:2004xq} 
are all what is needed for phenomenology for several years to come.
Although this is a legitimate claim, we believe that universal formalisms 
are interesting in themselves, and that their achievement should be pursued
in parallel to, and independently of, that of process-specific computations. 
Work in this direction is currently being performed by various groups,
and partial results are becoming 
available~\cite{Weinzierl:2003fx}--\cite{Somogyi:2005xz}.
We stress that the formulation of universal subtraction formalisms at NNLO
would pave the way to the matching between fixed-order computations and
parton-shower simulations, similarly to what recently done at
NLO~\cite{Frixione:2002ik,Frixione:2003ei}, thus resulting in predictions 
with a much broader range of applicability for phenomenological studies.

The purpose of the present paper is to propose a general framework for the
implementation of an observable- and process-independent subtraction method.
A complete formalism would require the construction of {\em all} of the NNLO
counterterms necessary to cancel soft and collinear singularities in the
intermediate steps of the calculation, as well as their integration over the
corresponding phase spaces. Most of the recent work on the subject
concentrated on the former aspect, without performing the
analytical integration of the kernels. In this paper we take a
different attitude: we propose a subtraction formula, and we give a
general prescription for the construction of all of the counterterms.
However, we do not construct most of them explicitly here, 
since we limit ourselves to considering only those
relevant to the $\CF\TR$ part of the dijet cross section in $e^+e^-$
collisions. On the other hand, we specifically address the problem of their
integration. We propose general formulae for the phase spaces necessary for
the integration of the subtraction kernels, and we use them to integrate the
counterterms mentioned above. In this way, we directly prove that the
subtraction formula we propose does allow us to cancel the singularities
at least in the simple case of the $\CF\TR$ contribution to the
$e^+e^-\to 2~{\rm jets}$ cross section, and we construct a numerical code with
which we recover the known results for dijet and total rates. Since the
subtraction formula, the counterterms, and the phase-space measures are
introduced in a way which is fully independent of the hard process considered,
this result gives us confidence that the framework we propose here is general
enough to lead to the formulation of a complete subtraction formalism, whose
explicit construction is, however, beyond the scope of the present paper.

The paper is organized as follows: in sect.~\ref{sec:NLO} we review
the strategy adopted by NLO universal subtraction formalisms,
and we formulate it in language suited to its extension to NNLO.
Such an extension is discussed in sect.~\ref{sec:NNLO}, and our 
main NNLO subtraction formula is introduced there. This subtraction 
procedure is shown to work in a simple case in sect.~\ref{sec:CFTR}. 
We present a short discussion in sect.~\ref{sec:comm}, and give
our conclusions in sect.~\ref{sec:concl}. A few useful formulae are 
collected in appendices~\ref{sec:simpl} and~\ref{sec:phsp}.

\section{Anatomy of subtraction at NLO\label{sec:NLO}}
Let us denote by $r$ any real (tree-level) matrix element squared 
contributing to the NLO correction of a given process, possibly times
a measurement function that defines an infrared-safe observable,
and by $d\psp$ the phase space. For example, when considering two-jet
production in $\epem$ collisions, $r$ is the product of the 
$\gamma^*\to q\bq g$ matrix element squared times the $\delta$ functions
that serve to define the jets within a given jet-finding algorithm,
and $d\psp$ is the three-body phase-space for the final-state partons 
$q$, $\bq$, and $g$. As is well known, the integral
\beq
R=\int r d\psp
\label{Rcont}
\eeq
is in general impossible to compute analytically. In the context of
the subtraction method, one rewrites eq.~(\ref{Rcont}) as follows:
\beq
R=\int\left(r d\psp - \tilde{r}d\tilde{\psp}\right)
+\int\tilde{r}d\tilde{\psp}\,.
\label{subtNLO}
\eeq
The quantities $\tilde{r}$ and $d\tilde{\psp}$ are completely arbitrary, 
except for the fact that $\tilde{r}d\tilde{\psp}$ must fulfill two conditions: 
the first integral in eq.~(\ref{subtNLO}) must be finite, and the second 
integral must be calculable analytically; when this happens, 
$\tilde{r}d\tilde{\psp}$ is called a {\em subtraction counterterm}.
The former condition guarantees that
the divergences of $\int\!\tilde{r}d\tilde{\psp}$ are the same as those
of $\int\! r d\psp$. Thus, by computing $\int\!\tilde{r}d\tilde{\psp}$ one 
is able to cancel explicitly, without any numerical inaccuracies, the 
divergences of the one-loop contribution, as stated by the KLN 
theorem; when using dimensional regularization with 
\beq
d=4-2\vep
\label{ddim}
\eeq
the divergences will appear as poles $1/\vep^k$, with $k=1,2$.
As far as the first integral in eq.~(\ref{subtNLO}) is concerned,
its computation is still unfeasible analytically. However, being
finite, one is allowed to remove the regularization, by letting
$\vep\to 0$, and to compute it numerically.

In the context of the subtraction method, the computation of an
observable to NLO accuracy therefore amounts to finding suitable forms 
for $\tilde{r}$ and $d\tilde{\psp}$, that fulfill the conditions given
above. Clearly, it is the asymptotic behaviour of $r$ in the soft
and collinear configurations that dictates the form of $\tilde{r}$.
The major difficulty here is that soft and collinear singularities
overlap, and $\tilde{r}$ should not oversubtract them. To show how
to construct $\tilde{r}$ systematically, let us introduce a few
notations. We call {\em singular limits} those configurations of 
four-momenta which may lead to divergences of the real matrix elements.
Whether the matrix elements actually diverge in a given limit depends 
on the identities of the partons whose momenta are involved in the limit; 
for example, no singularities are generated when a quark is soft, or 
when a quark and an antiquark of different flavours are collinear.
We call {\em singular (partonic) configurations} the singular limits 
associated with given sets of partons, and we denote them as follows:
\begin{itemize}
\item[$\bullet$] Soft: one of the partons has vanishing energy.
\beq
S_i:\;\;\;\;\;\;p_i^0\to 0.
\label{sNLO}
\eeq
\item[$\bullet$] Collinear: two partons have parallel three-momenta.
\beq
C_{ij}:\;\;\;\;\;\;\vp_i\parallel\vp_j.
\label{cNLO}
\eeq
\item[$\bullet$] Soft-collinear: two partons have parallel three-momenta,
and one of them is also soft.
\beq
SC_{ij}:\;\;\;\;\;\;(p_i^0\to 0,\vp_i\parallel\vp_j).
\label{scNLO}
\eeq
\end{itemize}
The singular limits will be denoted by $S$, $C$, and $SC$, i.e. by 
simply removing the parton labels in eqs.~(\ref{sNLO})--(\ref{scNLO}).
In practice, in what follows parton labels will be often understood, 
and singular limits will collectively indicate the corresponding 
partonic configurations.

For any function $f$, which depends on a collection of four-momenta, 
we introduce the following operator:
\beq
\ENLO(f)=f-\sum_i f(S_i)-\sum_{i<j}f(C_{ij})-\sum_{ij}f(SC_{ij}),
\label{ENLOdef}
\eeq
where $f(L_\alpha)$, $L_\alpha=S_\alpha,C_\alpha,SC_\alpha$ is the 
asymptotic behaviour of $f$ in the singular partonic 
configuration $L_\alpha$. Notice that 
the role of indices $i$ and $j$ is symmetric in the collinear limit,
but is not symmetric in the soft-collinear one (see eqs.~(\ref{cNLO})
and~(\ref{scNLO})), and this is reflected in the sums appearing in
the third and fourth terms on the r.h.s. of eq.~(\ref{ENLOdef}).
The asymptotic behaviour $f(L_\alpha)$ is always defined 
up to non-singular terms; however, what follows is independent
of the definition adopted for such terms. Although all of the singularities 
of $f$ are subtracted on the r.h.s. of eq.~(\ref{ENLOdef}), $\ENLO(f)$ is
not finite, owing to the overlap of the divergences. To get rid of this
overlap, we introduce a set of formal rules, that we call the
$\E$-{\em prescription}:
\begin{enumerate}
\item Apply $\ENLO$ to $f$, getting 
\mbox{$f+\sum_{\alpha_1} f_{\alpha_1}^{(1)}$}.
\item Apply $\ENLO$ to every $f_{\alpha_1}^{(1)}$ obtained in this way 
and substitute the result in the previous expression, 
getting \mbox{$f+\sum_{\alpha_1} f_{\alpha_1}^{(1)}
+\sum_{\alpha_2} f_{\alpha_2}^{(2)}$}.
\item Iterate the procedure until 
\mbox{$\ENLO(f_{\alpha_n}^{(n)})=f_{\alpha_n}^{(n)}$}, $\forall\,\alpha_n$,
 at fixed $n$.
\item Define \mbox{$\tilde{f}=-\sum_i\sum_{\alpha_i} f_{\alpha_i}^{(i)}$}.
\end{enumerate}
In order to show explicitly how the $\E$-prescription works, let us
apply it step by step. After the first iteration, we find eq.~(\ref{ENLOdef}).
With the second iteration, we need to compute
\beqn
\ENLO(f(S_i))&=&f(S_i)-\sum_{j\ne i} f(S_i\oplus S_j)-
\sum_{j<k}f(S_i\oplus C_{jk})-
\sum_{jk}^{j\ne i}f(S_i\oplus SC_{jk}),
\phantom{aaa}
\label{Etone}
\\
\ENLO(f(C_{ij}))&=&f(C_{ij})-\sum_k f(C_{ij}\oplus S_k)-
\sum_{k<l}^{\{i,j\}\ne\{k,l\}}f(C_{ij}\oplus C_{kl})
\nonumber\\*&&
-\sum_{kl}^{\{i,j\}\ne\{k,l\}}f(C_{ij}\oplus SC_{kl}),
\label{Ettwo}
\\
\ENLO(f(SC_{ij}))&=&f(SC_{ij})-\sum_k^{k\ne i} f(SC_{ij}\oplus S_k)-
\sum_{k<l}^{\{i,j\}\ne\{k,l\}}f(SC_{ij}\oplus C_{kl})
\nonumber\\*&&
-\sum_{kl}^{\{i,j\}\ne\{k,l\}}f(SC_{ij}\oplus SC_{kl}),
\label{Etthree}
\eeqn
where we denoted by $f(L_{\alpha_1}\oplus L_{\alpha_2})$ the asymptotic 
behaviour of the function $f(L_{\alpha_1})$ in the singular partonic
configuration $L_{\alpha_2}$. Although in general the operation $\oplus$ 
is non commutative, we shall soon encounter examples in which 
$f(L_{\alpha_1}\oplus L_{\alpha_2})=f(L_{\alpha_2}\oplus L_{\alpha_1})$.
Notice that we included all possible singular parton configurations 
in eqs.~(\ref{Etone})--(\ref{Etthree}), except for the redundant ones -- 
an example of which would be the case $k=i$ in the second term on the
r.h.s. of eq.~(\ref{Etthree}).

We now have to take into account the fact that we are performing a 
computation to NLO accuracy. Thus, the definition of an observable
will eventually be encountered (for example, embedded in a measurement
function), which will kill all matrix element singularities associated
with a partonic configuration that cannot contribute to the observable 
definition at NLO. For example, the limit in which two partons are soft 
is relevant only to beyond-NLO results, and this allows us to set
$f(S_i\oplus S_j)=0$ $\forall \{i,j\}$; analogously
\beq
\sum_k f(C_{ij}\oplus S_k)=
f(C_{ij}\oplus S_i)\,,
\eeq 
since the case $k\ne i$ would result into two unresolved partons,
which is again a configuration that cannot contribute to NLO.
In general, it is easy to realize that the operator $\ENLO$ is 
equivalent to the identity when it acts on the terms generated in the 
second iteration of the $\E$-prescription (i.e., the terms with negative
signs in eqs.~(\ref{Etone})--(\ref{Etthree})), and thus that, according
to the condition in item 3 above, the $\E$-prescription requires at most 
two iterations at NLO (it should be stressed that this would not be true
in the case of an infrared-unsafe observable, which would lead to an
infinite number of iterations). It follows that
\beq
\tilde{f}=\sum_i f(S_i)-\sum_{ij} f(S_i\oplus C_{ij})+
\sum_{i<j} f(C_{ij})-\sum_{ij} f(C_{ij}\oplus S_i)+\sum_{ij} f(SC_{ij}).
\eeq
This expression can be further simplified by observing that the soft 
and the collinear limits commute. This allows one to write
\beq
f(S_i\oplus C_{ij})=f(C_{ij}\oplus S_i)=f(SC_{ij}),
\label{commNLO}
\eeq
and therefore
\beq
\tilde{f}=\sum_i f(S_i)+\sum_{i<j}f(C_{ij})-\sum_{ij}f(SC_{ij}).
\label{ftilNLO}
\eeq
Let us now identify $f$ with $rd\psp$. Eq.~(\ref{ftilNLO}) implies that
the term added and subtracted in eq.~(\ref{subtNLO}) reads
\beq
\tilde{r}d\tilde{\psp}=\sum_i r(S_i)d\psp(S_i)+
\sum_{i<j} r(C_{ij})d\psp(C_{ij})-
\sum_{ij} r(SC_{ij})d\psp(SC_{ij}).
\label{rdmuNLO}
\eeq
All that is needed for the construction of a subtraction counterterm
at NLO is eq.~(\ref{rdmuNLO}), and the definition of the rules for 
the computations of $r(L)$ and $d\psp(L)$. In other words, all subtraction
procedures at NLO are implementation of eq.~(\ref{rdmuNLO}), i.e. of 
the $\E$-prescription, within a computation scheme for the asymptotic
behaviours of the matrix elements and the phase spaces. Although not
strictly necessary in principle, it is always convenient to adopt the
same procedure for the computation of $r(L)$ and of $d\psp(L)$; most 
conveniently, this is done by first choosing a parametrization for 
the phase space, and by eventually using it to obtain $r(L)$.

%%%%%%%%%%%%%%%%%%%%%%%%%%%%%%%%%%%%%%%%%%%%%%%%%%%%%%%%%%%%%%%%%%%
\begin{figure}[htb]
  \begin{center}
      \epsfig{figure=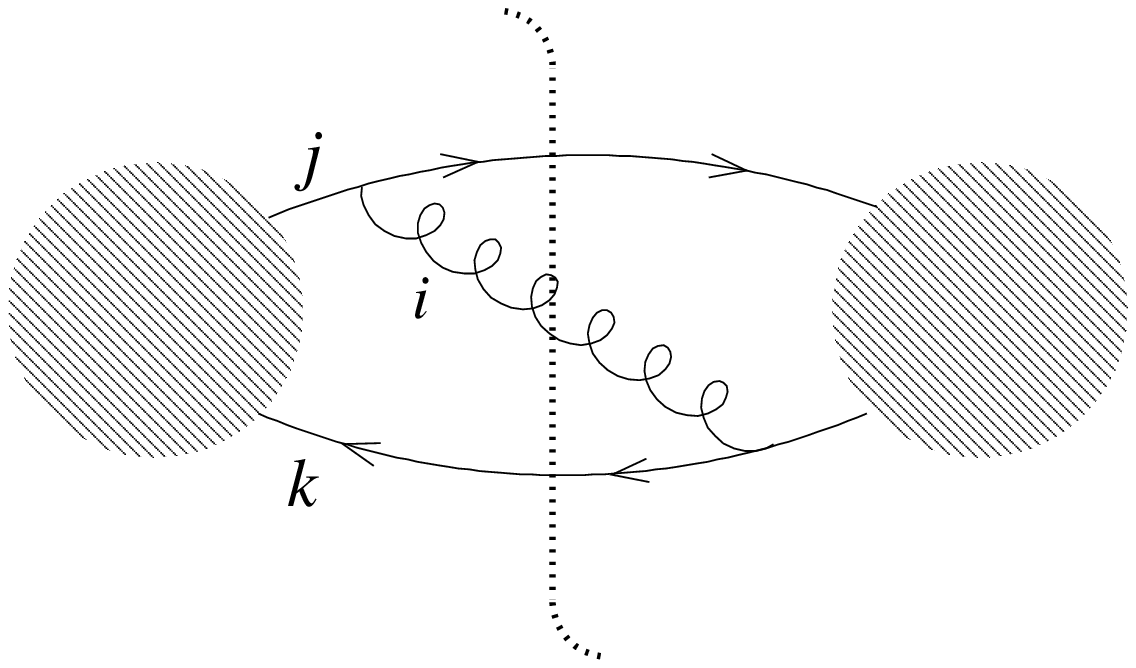,width=0.40\textwidth}
~~~~~~\epsfig{figure=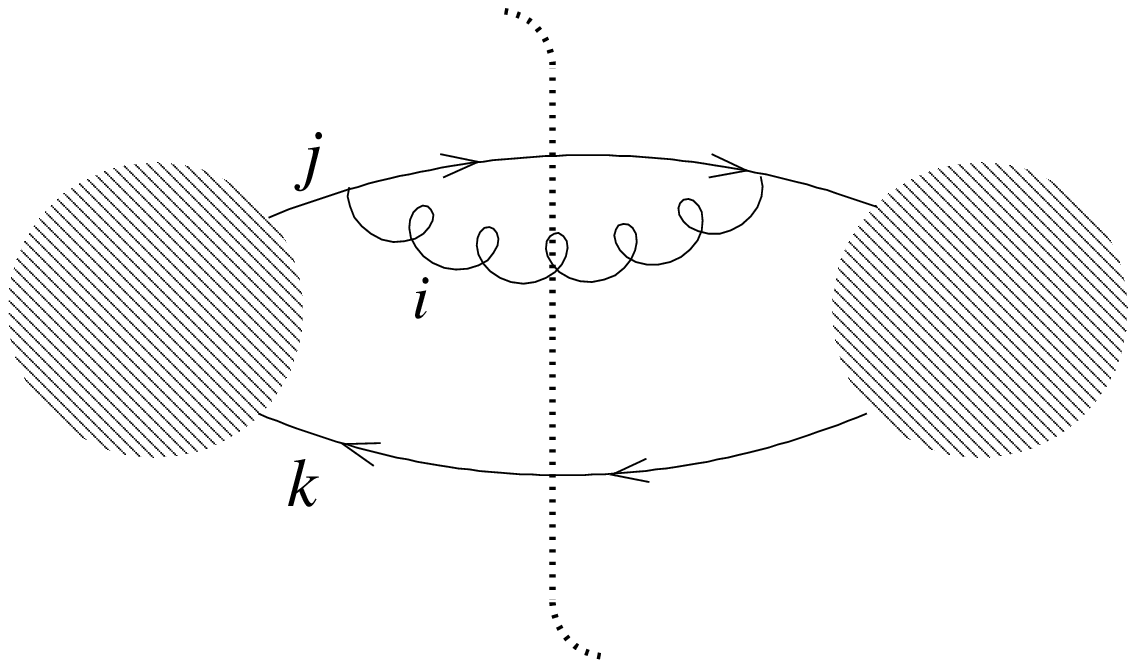,width=0.40\textwidth}
\caption{\label{fig:limits} 
Configurations contributing to soft (left panel)
and collinear (right panel) limits of squared amplitudes.  
}
  \end{center}
\end{figure}
%%%%%%%%%%%%%%%%%%%%%%%%%%%%%%%%%%%%%%%%%%%%%%%%%%%%%%%%%%%%%%%%%%%
In order to be more specific, we shall consider two explicit constructions
of subtraction counterterms, namely those of 
ref.~\cite{Frixione:1995ms} and of 
ref.~\cite{Catani:1996vz}. Since the subtractions always need to be
performed at the level of amplitudes squared, the relevant diagrams
(in a physical gauge) are those depicted in fig.~\ref{fig:limits},
for the soft (left panel) and collinear (right panel) limits
respectively. According to the notations introduced before and
the parton labelings that appear in the diagrams, we denote the
corresponding singular partonic configurations by $S_i$ and $C_{ij}$.
As is well known, the leading (singular) behaviours of the real matrix 
elements squared will be given by the following factors 
(see e.g. ref.~\cite{Bassetto:1984ik})
\beqn
&&S_i\;\longrightarrow\;\frac{p_j\mydot p_k}{p_i\mydot p_j p_i\mydot p_k}\,,
\label{softfc}
\\
&&C_{ij}\;\longrightarrow\;\frac{1}{p_i\mydot p_j}\,,
\label{collfc}
\eeqn
with the parton $k$ playing no role in the collinear limit.
In order to integrate the subtraction counterterms analytically,
a phase space parametrization must be chosen such that the leading
divergences displayed in eqs.~(\ref{softfc}) and~(\ref{collfc})
have as trivial as possible a dependence upon the integration
variables. Furthermore, the eikonal and collinear factors of
eqs.~(\ref{softfc}) and~(\ref{collfc}) have manifestly overlapping
divergences; thus, a matching treatment of the two, while not 
strictly necessary, would allow an easy identification of the 
overlapping contributions.

In ref.~\cite{Frixione:1995ms} the phase space is first decomposed 
in a manner which is largely arbitrary, but such that in each of 
the resulting regions only one soft and one collinear
singularity at most can arise (i.e., the other singularities are
damped by the $\stepf$ functions which are used to achieve 
the partition); the two may also occur simultaneously.
Thus, each region is identified by a pair of parton indices -- say,
$i$ and $j$ -- and no singularity other than \mbox{$1/p_i\mydot p_j$}
can occur in that region. This implies that the eikonal factor in
eq.~(\ref{softfc}) will not contribute a divergence to the region
above when $\vp_i\parallel\vp_k$.
%%%%%%%%%%%%%%%%%%%%%%%%%%%%%%%%%%%%%%%%%%%%%%%%%%%%%%%%%%%%%%%%%%%
\FIGURE[htb]{
      \epsfig{figure=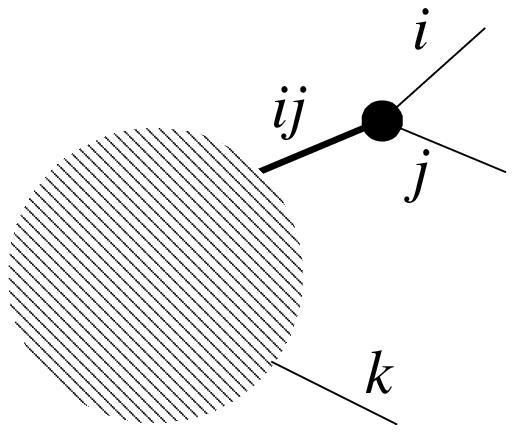,width=0.30\textwidth}
\caption{\label{fig:topone} 
Splitting topology contributing to NLO computations.
}}
%%%%%%%%%%%%%%%%%%%%%%%%%%%%%%%%%%%%%%%%%%%%%%%%%%%%%%%%%%%%%%%%%%%
In turn, this makes possible to choose a parametrization of the phase space, 
based on {\em exact} factorization formulae (see app.~\ref{sec:phsp}),
in which a pseudo-parton $i\!j$ (thick line in fig.~\ref{fig:topone}) 
branches into on-shell partons $i$ and $j$ (narrow lines); in other
words, the phase space of eq.~(\ref{Rcont}) is written as follows:
\beq
d\psp = d\psp_{n-1}(i\!j)d\pspb_2(i,j)\,,
\label{psfactFKS}
\eeq
where $d\pspb$ is the two-body phase space, times the measure over
the virtuality of the branching pseudo-parton $i\!j$. 
The parton $k$ may or may not serve to define the integration variables, 
but is irrelevant in the treatment of the singularities. Clearly, this 
parametrization is suggested by a collinear-like configuration, 
but thanks to the partition of the phase space it also allows a 
straightforward integration over soft singularities. Graphically,
this is equivalent to squaring the parts of the diagrams in 
fig.~\ref{fig:limits} which lay to the left of the cut. The contribution
due to the part to the right of the cut in the diagram on the left panel
is clearly recovered once the sum over parton labels is carried out,
since the role of indices $j$ and $k$ is fully symmetric. Once the
exact parametrization of eq.~(\ref{psfactFKS}) is fixed, 
ref.~\cite{Frixione:1995ms} proceeds by defining
\beq
d\psp(L) = d\psp_{n-1}(\widetilde{i\!j})d\pspb_2(i,j;L)\,,
\label{psfactFKSs}
\eeq
where $\widetilde{i\!j}$ is an on-shell parton, and
\beqn
d\pspb_2(i,j;S_i)&=&d\pspb_2(i,j)\Big|_{p_i^0\to 0}\,,
\\
d\pspb_2(i,j;C_{ij})&=&d\pspb_2(i,j)\Big|_{\vp_i\parallel\vp_j}\,,
\\
d\pspb_2(i,j;SC_{ij})&=&d\pspb_2(i,j)\Big|_{p_i^0\to 0,\vp_i\parallel\vp_j}\,,
\eeqn
are obtained by neglecting constant terms in the corresponding limits.
Obviously, $d\psp(L)$ is not an exact representation of the full phase space
any longer, i.e. $d\psp(L)\ne d\psp$; however, this does not introduce 
any approximation in the procedure, since the subtraction counterterm is 
subtracted and added back in the physical cross section (eq.~(\ref{subtNLO})).
Finally, the asymptotic behaviours $r(L)$ appearing in eq.~(\ref{rdmuNLO})
are directly taken from factorization formulae, eqs.~(\ref{softfc}) 
and~(\ref{collfc}), without any further manipulation.

In the dipole formalism of ref.~\cite{Catani:1996vz}, the
following identity is exploited 
\beq
\frac{1}{p_i\mydot p_j p_i\mydot p_k}=
\frac{1}{p_i\mydot p_j p_i\mydot (p_j+p_k)}+
\frac{1}{p_i\mydot p_k p_i\mydot (p_j+p_k)}
\label{eikid}
\eeq
for the eikonal factor of eq.~(\ref{softfc}).
The two terms on the r.h.s. of eq.~(\ref{eikid}) are symmetric
for $j\leftrightarrow k$, and thus only the first
one actually needs to be considered.
As far as collinear configurations are concerned, this term is singular 
only when $\vp_i\parallel\vp_j$, but not when $\vp_i\parallel\vp_k$.
Thus, the identity in eq.~(\ref{eikid}) has the same function as the 
partition of the phase space of ref.~\cite{Frixione:1995ms}.
Furthermore, an exact parametrization of the phase space is chosen
\beq
d\psp = d\psp_{n-1}(\widetilde{i\!j};k)d\pspb_2(i,j;k)
\label{psfactCS}
\eeq
which differs from the one of eq.~(\ref{psfactFKS}) in that the parton
$k$ (called the spectator) plays a fundamental role, since it allows
to put on shell the splitting pseudo-parton even if $i$ and $j$ are
not exactly collinear, or $i$ is not soft. Thanks to this property,
in ref.~\cite{Catani:1996vz} we have
\beq
d\psp(L) = d\psp\,,\;\;\;\;\;\;\;\;
L=S_i,C_{ij},SC_{ij},
\eeq
and thus eq.~(\ref{rdmuNLO}) becomes
\beqn
&&\tilde{r}d\tilde{\psp}=D_{ij;k} d\psp\,,
\label{dipsubt}
\\*
&&D_{ij;k}=r(S_{i;k})+r(C_{ij})-r(SC_{ij})\,,
\label{NLOdip}
\eeqn
where $r(S_{i;k})$ is obtained by taking the soft limit of the real 
matrix element, and keeping only the first term on the r.h.s. of 
eq.~(\ref{eikid}).

In summary, the common feature of 
refs.~\cite{Frixione:1995ms,Catani:1996vz} is the fact
that, by disentangling (with different techniques) the two collinear
singularities that appear in each eikonal factor, they define subtraction
formalisms based on building blocks which all have a collinear-like topology;
we shall denote by $\To$ this topology, depicted in fig.~\ref{fig:topone}
(the parton labels are obviously irrelevant for topological considerations),
a notation which is reminiscent of a branching after which the list of 
resolved partons is diminished by one unity.
The explicit expressions for the building blocks, which originate from
the soft, collinear, and soft-collinear limits and include the treatment
of the phase space, depend on the formalism; however, in all
cases they are combined according to eq.~(\ref{rdmuNLO}), i.e. according 
to the $\E$-prescription, in order to construct the sought subtraction
counterterm.

\section{Subtraction at NNLO\label{sec:NNLO}}
In this section, we shall introduce a general framework for the
implementation of a subtraction method to NNLO accuracy. 
We shall consider the process
\beq
\epem\,\longrightarrow\,n~{\rm jets}\,.
\label{physproc}
\eeq
In this way, all the intricacies are avoided due to initial-state
collinear singularities, which allows us to simplify the notation 
considerably. The systematic construction of the subtraction counterterms
that we propose in the following will however be valid also in the 
case of processes with initial-state hadrons, since the procedure is
performed at the level of short-distance partonic cross sections.
On the other hand, we do not present here the explicit parametrizations 
of the phase spaces for the case of initial-state partons, and we do
not consider the contributions of initial-state collinear counterterms
%arising from the bare parton distribution functions,
which are necessary
in order to achieve the complete cancellation of infrared singularities
for processes with QCD partons in the initial state.

\subsection{Generalities\label{sec:NNLOintro}}
At NNLO, the process~(\ref{physproc}) receives contribution from the 
following partonic subprocesses
\beq
\epem\,\longrightarrow\,m~{\rm partons},\;\;\;\;\;\;
m=n,\,n+1,\,n+2.
\label{partproc}
\eeq
We write the amplitude corresponding to eq.~(\ref{partproc}) in the
following way
\beq
\Am=\gs^{m-2}\Tm + \gs^m \LOm + \gs^{m+2} \LTm+\dots,
\label{ampm}
\eeq
where $\Tm$, $\LOm$ and $\LTm$ are the tree-level, one-loop and
two-loop contributions to the process~(\ref{partproc}) respectively. 
Squaring eq.~(\ref{ampm}) we get:
\beqn
\abs{\An}^2&=&\gs^{2n-4}\abs{\Tn}^2
+\gs^{2n-2}\left(\Tn\LOnStar+\TnStar\LOn\right)
\nonumber \\*&&
+\gs^{2n}\left(\abs{\LOn}^2+\Tn\LTnStar+\TnStar\LTn\right)
+{\cal O}(\gs^{2n+2}),
\label{ampsqn}
\\
\abs{\Anone}^2&=&\gs^{2n-2}\abs{\Tnone}^2
+\gs^{2n}\left(\Tnone\LOnoneStar+\TnoneStar\LOnone\right)
+{\cal O}(\gs^{2n+2}),
\label{ampsqnone}
\\
\abs{\Antwo}^2&=&\gs^{2n}\abs{\Tntwo}^2
+{\cal O}(\gs^{2n+2}).
\label{ampsqntwo}
\eeqn
In eq.~(\ref{ampsqn}) the number of partons coincides with the number
of jets of the physical process, and this implies that all partons
must be {\em resolved}, i.e. hard and well separated. On the other
hand, in eqs.~(\ref{ampsqnone}) and~(\ref{ampsqntwo}) the number of
partons exceeds that of jets, which means that one and two partons
respectively are unresolved in these contributions. The accuracy
with which the various terms in eqs.~(\ref{ampsqn})--(\ref{ampsqntwo}) 
enter the cross section can be read from the power of $\gs$.
The Born contribution is proportional to $\gs^{2n-4}$, and appears
solely in eq.~(\ref{ampsqn}). The NLO contributions are proportional to 
$\gs^{2n-2}$, and appear in eqs.~(\ref{ampsqn}) and~(\ref{ampsqnone}).
The divergences of the former are entirely due to the loop integration
implicit in $\LOn$, whereas those of the latter are obtained analytically 
after applying the subtraction procedure described in sect.~\ref{sec:NLO}
(there, $\gs^{2n-2}\abs{\Tnone}^2$ has been denoted by $r$). We understand
that, in the actual computation of an infrared-safe observable, the
matrix elements in eqs.~(\ref{ampsqn})--(\ref{ampsqntwo}) are multiplied
by the relevant measurement functions.

The NNLO contributions are proportional to $\gs^{2n}$, and we can
classify them according to the number of unresolved partons. 
In the {\em double-virtual} contribution, 
$\abs{\LOn}^2+(\Tn\LTnStar+\TnStar\LTn)$, all partons are resolved. 
The term in round brackets is identical to the NLO virtual contribution,
except for the fact that one-loop results are formally replaced 
by two-loop ones; on the other hand, the former term is typologically 
new. However, for both the structure of the singularities is explicit
once the loop computations are carried out. We then have the 
{\em real-virtual} contribution, $\Tnone\LOnoneStar+\TnoneStar\LOnone$,
in which one parton is unresolved. This is again formally identical
to the NLO virtual contribution, but there is a substantial difference:
in addition to the singularities resulting from loop integration,
there are singularities due to the unresolved parton, which will appear 
explicitly only after carrying out the integration over its phase space.
In order to do this analytically, a subtraction procedure will be 
necessary, and the methods of sect.~\ref{sec:NLO} may be applied.
When doing so, however, at variance with an NLO computation the analogue 
of the first term on the r.h.s. of eq.~(\ref{subtNLO}) will not be finite, 
because of the presence of the explicit divergences due to the one-loop
integration. This prevents us from setting $\vep\to 0$ as in  
eq.~(\ref{subtNLO}), and thus ultimately from computing
the integral, since this integration can only be done numerically.
It follows that the straightforward application of an NLO-type
subtraction procedure to the real-virtual contribution {\em alone} 
would not lead to the analytical cancellation of all the divergences.

We shall show that such a cancellation can be achieved by adding to the
real-virtual contribution a set of suitably-defined terms obtained from 
the {\em double-real} contribution, $\abs{\Tntwo}^2$. This contribution 
is characterized by the fact that two partons are unresolved and, analogously
to the case of the real contribution to an NLO cross section, all of the
divergences are obtained upon phase space integration, a task which is overly
complicated due to the substantial amount of overlapping among the various
singular limits.

\subsection{The subtracted cross section\label{sec:NNLOrr}}
In this section we shall use the findings of sect.~\ref{sec:NLO} as a template 
for the systematic subtraction of the phase-space singularities of the
double-real contribution, which we shall obtain by suitably generalize 
the $\E$-prescription. In order to do this, we start from listing all 
singular limits that lead to a divergence of the double-real matrix
elements. We observe that the $S_\alpha$, $C_\alpha$, and $SC_\alpha$ 
configurations described in sect.~\ref{sec:NLO} are also relevant to 
the double-real case. In addition, we have the following 
configurations~\cite{Campbell:1997hg,Catani:1999ss}:
\begin{itemize}
\item[$\bullet$] Soft-collinear: two partons have parallel three-momenta,
and a third parton has vanishing energy\footnote{Note therefore that,
upon removing the parton labels, the $SC$ symbol denotes both the cases 
with one or two unresolved partons, eqs.~(\ref{scNLO}) and~(\ref{IRSCt})
respectively.}.
\beq
SC_{ijk}:\;\;\;\;\;\;(p_i^0\to 0,\vp_j\parallel\vp_k).
\label{IRSCt}
\eeq
\item[$\bullet$] Double soft: two partons have vanishing energy.
\beq
SS_{ij}:\;\;\;\;\;\;(p_i^0\to 0,p_j^0\to 0).
\eeq
\item[$\bullet$] Double collinear: three partons have parallel three-momenta,
or two pairs of two partons have parallel three-momenta\footnote{The
former case is usually denoted as triple collinear. We prefer this 
notation since, if $\vec{a}\parallel\vec{b}$ and $\vec{b}\parallel\vec{c}$, 
necessarily $\vec{a}\parallel\vec{c}$. Furthermore, the present notation is 
more consistent with the strongly-ordered limit case. It gives minimal but
sufficient information.}.
\beqn
&&CC_{ijk}:\;\;\;\;\;\;(\vp_i\parallel\vp_j\parallel\vp_k),
\label{IRCCt}
\\*
&&CC_{ijkl}:\;\;\;\;\;\;(\vp_i\parallel\vp_j,\vp_k\parallel\vp_l).
\label{IRCCopo}
\eeqn
\item[$\bullet$] Double soft and collinear: two partons have vanishing energy,
and two partons have parallel three-momenta.
\beqn
&&SSC_{ij}:\;\;\;\;\;\;(p_i^0\to 0,p_j^0\to 0,\vp_i\parallel\vp_j),
\\*
&&SSC_{ijk}:\;\;\;\;\;\;(p_i^0\to 0,p_j^0\to 0,\vp_j\parallel\vp_k).
\eeqn
\item[$\bullet$] Double collinear and soft: as in the case of double
collinear, but one of the collinear partons has also vanishing energy.
\beqn
&&SCC_{ijk}:\;\;\;\;\;\;(p_i^0\to 0,\vp_i\parallel\vp_j\parallel\vp_k),
\\*
&&SCC_{ijkl}:\;\;\;\;\;\;(p_i^0\to 0,\vp_i\parallel\vp_j,\vp_k\parallel\vp_l).
\eeqn
\item[$\bullet$] Double soft and double collinear: as in the case of double
collinear, but two of the collinear partons have also vanishing energy.
\beqn
&&SSCC_{ijk}:\;\;\;\;\;\;
(p_i^0\to 0,p_j^0\to 0,\vp_i\parallel\vp_j\parallel\vp_k),
\\*
&&SSCC_{ijkl}:\;\;\;\;\;\;
(p_i^0\to 0,p_k^0\to 0,\vp_i\parallel\vp_j,\vp_k\parallel\vp_l).
\label{IRSSCCopo}
\eeqn
\end{itemize}
The soft and collinear limits in each of eqs.~(\ref{IRSCt})--(\ref{IRSSCCopo})
are understood to be taken simultaneously, following for example
the rules given in ref.~\cite{Catani:1999ss} (see in particular
eqs.~(23) and~(98) there). 

%%%%%%%%%%%%%%%%%%%%%%%%%%%%%%%%%%%%%%%%%%%%%%%%%%%%%%%%%%%%%%%%%%%
\begin{figure}[htb]
  \begin{center}
      \epsfig{figure=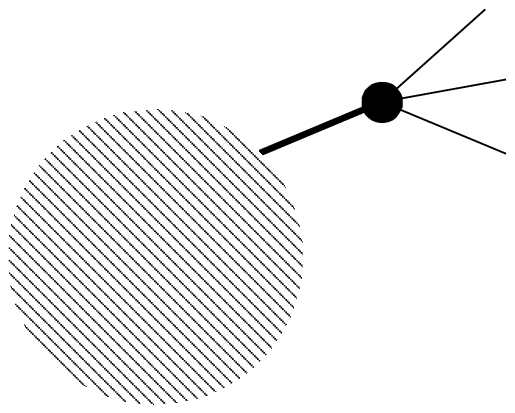,width=0.25\textwidth}
~~~~~~~~~~~~\epsfig{figure=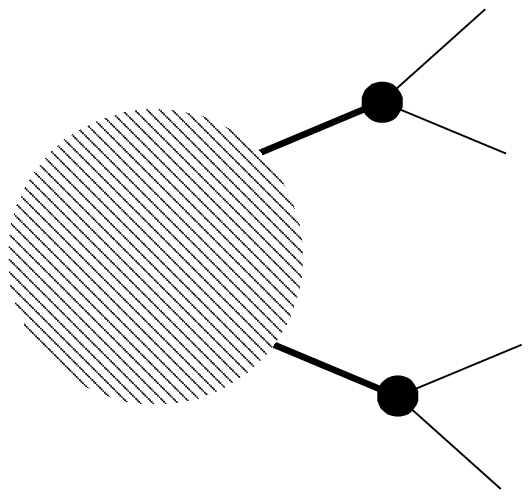,width=0.25\textwidth}
\caption{\label{fig:topNNLO} 
NNLO-type topologies: $\Tt$ (left panel) and $\Topo$ (right panel).
}
  \end{center}
\end{figure}
%%%%%%%%%%%%%%%%%%%%%%%%%%%%%%%%%%%%%%%%%%%%%%%%%%%%%%%%%%%%%%%%%%%
The necessity of introducing the notion of
topology emerges at NNLO even without considering the soft
limits and the problem of overlapping divergences, as is clear
by inspection of the purely collinear limits, eqs.~(\ref{IRCCt})
and~(\ref{IRCCopo}). These are associated with the
branching processes depicted in fig.~\ref{fig:topNNLO}; we denote
the corresponding topologies by $\Tt$ and $\Topo$ respectively
(again, this notation serves as a reminder of the number of 
partons to be removed from the list of resolved partons).
Some of the singular limits in which one or two partons are soft 
cannot be straightforwardly associated with either topology. However, 
as in the case of NLO computations, this can be done after 
some formal manipulations, whose nature (be either a partition
of the phase space, or a partial fractioning, or something else) we do 
not need to specify at this stage. Suffice here to say that, after such
manipulations, all singular limits in eqs.~(\ref{IRSCt})--(\ref{IRSSCCopo})
that feature at least one soft parton will in general contribute to 
both $\Tt$ and $\Topo$ topologies (we may formally write, for 
example, $f(SS)=f(SS^{(\Tt)})+f(SS^{(\Topo)})$). The same kind 
of procedure can be applied to the singular limits of 
eqs.~(\ref{sNLO})--(\ref{scNLO}), since topology $\To$ can always
be seen as a sub-topology of $\Tt$ or $\Topo$; thus, $S$, $C$, and
$SC$ limits will be manipulated, if need be, so as to be associated
with topologies $\Tt$ and $\Topo$. In the case we shall need to
distinguish between the singular limits in the various topologies,
we shall denote them by $L^{(\toplab)}$, with $\toplab=\To,\Tt,\Topo$. 
However, as for parton labels, topology labels may be understood
in what follows.

We now claim that by applying the $\E$-prescription defined in 
sect.~\ref{sec:NLO} we can systematically subtract the singularities
of the double-real contribution to the NNLO cross section, provided
that the operator $\ENLO$ is replaced by $\ENNLO$, where
\beqn
\ENNLO(f)&=&\ENLO(f)-\sum_{\toplab=\Tt,\Topo}\sum_\alpha
\Bigg\{f\left(SC_\alpha^{(\toplab)}\right)+f\left(SS_\alpha^{(\toplab)}\right)+
f\left(CC_\alpha^{(\toplab)}\right)
\nonumber\\*&&\phantom{\sum_{\toplab=\Tt,\Topo}\sum_\alpha}
+f\left(SSC_\alpha^{(\toplab)}\right)+f\left(SCC_\alpha^{(\toplab)}\right)+
f\left(SSCC_\alpha^{(\toplab)}\right)\Bigg\},
\label{ENNLOdef}
\eeqn
and $\alpha$ denotes all indices relevant to the corresponding singular
limits, which can be read in eqs.~(\ref{IRCCt})--(\ref{IRSSCCopo}).
As in the case of NLO computations, the iterative $\E$-prescription 
comes to an end thanks to the infrared safety of the observables.
However, at NNLO up to four iterations are necessary in order to 
define $\tilde{f}$, which can be easily generated by means of an
algebraic-manipulation code. In such a way, one obtains up to 51 
terms for each topology; fortunately, such a massive counterterm
can be greatly simplified. We start by observing that the freedom
in the definition of the asymptotic form of the matrix elements
associated with a given singular limit allows us to exploit commutation 
properties as done in eq.~(\ref{commNLO}). Furthermore, the presence of
$\ENLO$ in the definition of $\ENNLO$ implies that some of the terms 
obtained with the $\E$-prescription will be formally identical to those 
appearing in an NLO subtraction. This suggests us to write
\beq
\tilde{f}=\tfmo+\tfmt^{(\Tt)}+\tfmt^{(\Topo)}\,,
\label{ftilNNLOt}
\eeq
where
\beq
\tfmo=\sum_i f\left(S_i^{(\To)}\right)+
\sum_{i<j}f\left(C_{ij}^{(\To)}\right)-
\sum_{ij}f\left(SC_{ij}^{(\To)}\right),
\label{tfmodef}
\eeq
and
\beqn
\tfmt^{(\toplab)}&=&\sum_\alpha\Bigg\{
f\left(CC_\alpha^{(\toplab)}\right)+f\left(SS_\alpha^{(\toplab)}\right)
-f\left(SC_\alpha^{(\toplab)}\right)
-f\left([C\oplus SS]_\alpha^{(\toplab)}\right)
\nonumber \\*&&\phantom{\sum_\alpha}
-f\left([CC\oplus S]_\alpha^{(\toplab)}\right)
-f\left([CC\oplus SS]_\alpha^{(\toplab)}\right)
-f\left([S\oplus S]_\alpha^{(\toplab)}\right)
-f\left([C\oplus C]_\alpha^{(\toplab)}\right)
\nonumber \\*&&\phantom{\sum_\alpha}
+f\left([C\oplus S\oplus S]_\alpha^{(\toplab)}\right)
+f\left([CC\oplus S\oplus S]_\alpha^{(\toplab)}\right)
+f\left([C\oplus C\oplus S]_\alpha^{(\toplab)}\right)
\nonumber \\*&&\phantom{\sum_\alpha}
+f\left([C\oplus C\oplus SS]_\alpha^{(\toplab)}\right)
-f\left([C\oplus C\oplus S\oplus S]_\alpha^{(\toplab)}\right)\Bigg\},
\label{tfmtdef}
\eeqn
where we used the fact that (for example) $SSC=SS\oplus C=C\oplus SS$, 
and $\alpha$ denotes symbolically the relevant parton indices,
not indicated explicitly in order to simplify the notation. The physical
meaning of eqs.~(\ref{ftilNNLOt})--(\ref{tfmtdef}) is clear: in the
double real contribution to an NNLO cross section, there are terms with
one or two unresolved partons. The former have the same kinematics as
those relevant to a pure NLO subtraction. More interestingly, they also
have the same kinematics as the real-virtual contribution. Although this 
fact formally results from the application of the $\E$-prescription, 
it can also be understood intuitively: by requiring more stringent 
jet-finding conditions (for example, by enlarging the minimum
$\pt$ which defines a tagged jet), the NNLO cross section turns
into an NLO one, which receives contributions only from the real-virtual
term, and from the pieces obtained by applying $\ENLO$ to the double-real
matrix elements. Clearly, if $f$ is associated with an $(n+2)$-body final 
state, each term in $\tfmo$ and $\tfmt$ factorize $(n+1)$-body and $n$-body 
hard matrix elements and measurement functions respectively.

In order to give more details on the structure of the subtraction that
emerges from the $\E$-prescription, let us denote by
\beqn
vv&=&\gs^{2n}\left(\abs{\LOn}^2+\Tn\LTnStar+\TnStar\LTn\right),
\\
rv&=&\gs^{2n}\left(\Tnone\LOnoneStar+\TnoneStar\LOnone\right),
\\
rr&=&\gs^{2n}\abs{\Tntwo}^2,
\eeqn
the double-virtual, real-virtual, and double-real matrix elements
squared respectively, possibly times measurement functions that 
we understand. The jet cross section is
\beqn
d\sigma&=&d\sigma_{rr}+d\sigma_{rv}+d\sigma_{vv}
\\*
&=&\int rr d\psp_{n+2}+\int rv d\psp_{n+1}+\int vv d\psp_n\,.
\label{jetxsec}
\eeqn
We start by applying the $\E$-prescription to the double-real contribution
\beq
d\sigma_{rr}=\int\left(rr d\psp_{n+2}-\trrmt d\tilde{\psp}_{n+2}^{-2}
-\trrmo d\tilde{\psp}_{n+2}^{-1}\right)
+\int\trrmt d\tilde{\psp}_{n+2}^{-2}
+\int\trrmo d\tilde{\psp}_{n+2}^{-1},
\label{rrsubt}
\eeq
where we allowed the possibility of adopting two different parametrizations
for the phase spaces attached to $\trrmt$ and to $\trrmo$.
The first integral on the r.h.s. of eq.~(\ref{rrsubt}) is finite, and can
be computed numerically after removing the regularization by letting
$\vep\to 0$. The second and the third integrals will contain all the 
divergences of the double-real contribution, to be cancelled by those 
of the real-virtual and double-virtual contributions. However, only the 
second term can, at this stage, be integrated analytically. In fact,
$\trrmo$ is, according to its definition, obtained by considering
the asymptotic behaviour of the double-real matrix element squared 
in the singular limits $S$, $C$, and $SC$ of NLO nature. These limits
will render manifest only part of the singular structure of the matrix
elements, preventing a complete analytical integration of the divergent
terms. For example, if parton $i$ becomes soft, $rr(S_i)$ factors the 
eikonal term of eq.~(\ref{softfc}), and the integration over the
variables of parton $i$ can be carried out analytically, as in the
case of NLO. However, at NNLO we must take into account that
another singularity may appear -- say, parton $l$ may also become
soft, and $rr(S_i)$ has too complicated a dependence upon the variables
of parton $l$ to perform the necessary analytical integration. Notice that
this is not true for, say, $rr(SS_{il})$, which is why the second term on the 
r.h.s. of eq.~(\ref{rrsubt}) can indeed be integrated. This suggests 
combining $\trrmo$ with the real-virtual contribution which, as discussed in
sect.~\ref{sec:NNLOintro}, cannot be integrated analytically too
(although for different reasons). In order to achieve this combination,
we must choose the phase-space measure appropriately; in particular, 
we shall use
\beq
d\tilde{\psp}_{n+2}^{-1}=
d\psp_{n+1}d\psp_2^{-1}\,,
\label{rvphsp}
\eeq
where $d\psp_{n+1}$ is the exact $(n+1)$-body massless phase space, 
and $d\psp_2^{-1}$ is related to the phase space relevant to the partons 
whose contributions to the singular behaviour of the double-real
matrix elements is explicit in $\trrmo$, and includes the measure over 
the virtuality of the branching parton; we shall show in
what follows how to construct explicitly the phase spaces
of eq.~(\ref{rvphsp}). Having done that, we define a subtracted
real-virtual contribution as follows
\beq
rv^{(s)}=rv+\int\trrmo d\psp_2^{-1}\,.
\label{rvsubt}
\eeq
In an NLO computation, eq.~(\ref{rvsubt}) would amount to the
full NLO correction, with the first and the second term on the
r.h.s. playing the roles of virtual and real contributions respectively.
This implies that the {\em explicit} poles in $1/\vep$ that appear
in $rv$ because of the loop integration will be exactly cancelled by those 
of $\trrmo$ which result from the phase-space integration $d\psp_2^{-1}$.
Thus, although in an NNLO computation $rv^{(s)}$ still contains 
phase-space divergences, we can manipulate it in the same manner 
as a real contribution to an NLO cross section:
\beq
\int rv^{(s)}d\psp_{n+1}=
\int\left(rv^{(s)} d\psp_{n+1} - \widetilde{rv}^{(s)}d\tilde{\psp}_{n+1}\right)
+\int\widetilde{rv}^{(s)}d\tilde{\psp}_{n+1}\,.
\label{rvfinal}
\eeq
The last term on the r.h.s. of eq.~(\ref{rvfinal}) can now be integrated
analytically, whereas the first is finite and can be integrated numerically.
Therefore, by combining part of the double-real and the real-virtual
contributions we have managed to define a scheme in which the analytical
integration of all the divergent terms is possible. By combining 
eqs.~(\ref{jetxsec}), (\ref{rrsubt}), and~(\ref{rvfinal}) we get
\beqn
d\sigma&=&
\int\left(rr d\psp_{n+2}-\trrmt d\tilde{\psp}_{n+2}^{-2}
-\trrmo d\tilde{\psp}_{n+2}^{-1}\right)
\nonumber\\*
&+&\int\left(rv^{(s)} d\psp_{n+1}- 
\widetilde{rv}^{(s)}d\tilde{\psp}_{n+1}\right)
\nonumber\\*
&+&\int\trrmt d\tilde{\psp}_{n+2}^{-2}
+\int\widetilde{rv}^{(s)}d\tilde{\psp}_{n+1}
+\int vv d\psp_n\,,
\label{jetxsecsub}
\eeqn
where the first two terms on the r.h.s. are finite, and can be integrated
numerically after letting $\vep\to 0$; the sum of the remaining terms is
also finite, but they are individually divergent and must be computed
analytically.

In order to show explicitly how our master subtraction formula 
eq.~(\ref{jetxsecsub}) works, we consider the unphysical case in
which only singular limits of collinear nature can contribute to 
the cross section. In such a situation, the application of the
$\E$-prescription is trivial, and we readily arrive at\footnote{The
parton indices play an obvious role here, and we omit them in order
to simplify the notation.}:
\beqn
&&\trrmt\, d\tilde{\psp}_{n+2}^{-2}=
rr(CC)\,d\tilde{\psp}_{n+2}^{-2}(CC)-
rr(C\oplus C)\,d\tilde{\psp}_{n+2}^{-2}(C\oplus C)\,,
\label{unprrmt}
\\
&&\trrmo\, d\tilde{\psp}_{n+2}^{-1}=rr(C)\,d\tilde{\psp}_{n+2}^{-1}(C)\,.
\label{unprrmo}
\eeqn
As in the general formula for the subtraction counterterm at NLO,
eq.~(\ref{rdmuNLO}), we leave the possibility open of associating 
different phase spaces measures with different terms in $\trrmt$. This 
has the advantage that each parametrization can be tailored in order 
to simplify as much as possible the analytical integration. The
drawback is that, in general, there could be algebraic simplifications
among the contributions to $\trrmt$ (here, between $rr(CC)$ and
$rr(C\oplus C)$), which can be explicitly carried out only upon
factorizing the phase space. We shall discuss this point further
in the following, in the context of a more physical case.

By construction, $rr(C\oplus C)$ is the strongly-ordered, double-collinear
limit. This may coincide with $rr(CC)$, and it does in particular in
topology $\Topo$, but in general $\trrmt$ is different from 
zero\footnote{Here and in what follows, ``zero'' means non-divergent.}, 
and corresponds to the non-strongly-ordered part of the double-collinear
limit. As such, when a kinematic configuration is generated in which
two, and only two partons are collinear, we have $\trrmt\to 0$; in
other words, the $C$ limit of $\trrmt$ is zero. This is what should 
happen: in the collinear limit, $rr(C)$ (which appears in $\trrmo$) 
is sufficient to cancel locally the divergences of $rr$, and thus 
$\trrmt$ should not diverge in this limit, since otherwise the first 
term on the r.h.s. of eq.~(\ref{jetxsecsub}) would not be finite.
Analogously, in the double-collinear limit the local counterterm
for $rr$ is $rr(CC)$; therefore, in order to avoid divergences,
in such a limit the contribution of $rr(C\oplus C)$ (in $\trrmt$) 
must cancel that of $rr(C)$ (in $\trrmo$). Notice that, for these 
cancellations to happen not only at the level of matrix elements,
but also at the level of cross sections, suitable choices of the phase
spaces must be made. The $C$ limit thus relates $d\tilde{\psp}_{n+2}^{-2}(CC)$
to $d\tilde{\psp}_{n+2}^{-2}(C\oplus C)$, whereas the $CC$ limit relates
$d\tilde{\psp}_{n+2}^{-2}(C\oplus C)$ to $d\tilde{\psp}_{n+2}^{-1}(C)$. 
A suitable choice in the former case clearly includes the trivial one,
where the two phase spaces are taken to be identical.

We finally note that the subtraction formula of eq.~(\ref{jetxsecsub}) appears
in a very similar form in ref.~\cite{Weinzierl:2003fx}, where a first
discussion was given on the extension of the subtraction method to NNLO.
Ref.~\cite{Weinzierl:2003fx} constructs the NNLO subtraction formula
building upon the NLO dipole subtraction~\cite{Catani:1996vz}, and provides 
the explicit expressions of the NNLO counterterms for the leading-colour 
contribution to $e^+e^-\to 2~{\rm jets}$. Ref.~\cite{Weinzierl:2003fx},
however, does not discuss the way in which the subtraction kernels
can be constructed for more general processes. In our approach, the
$\E$-prescription provides a general framework for the construction of the
counterterms $\trrmo$, $\trrmt$ and $\widetilde{rv}^{(s)}$, and explicitly
suggests the subtractions of eq.~(\ref{jetxsecsub}). We find it reassuring
that we arrive at a subtraction structure consistent with that of
ref.~\cite{Weinzierl:2003fx}, given the fact that neither here nor in
ref.~\cite{Weinzierl:2003fx} a formal proof is given that
eq.~(\ref{jetxsecsub}) achieves a complete cancellation of the infrared
singularities.  We do obtain such a cancellation explicitly, for the $\CF\TR$
colour factor of the dijet cross section, as we shall show in the next
section. On the other hand, to the best of our knowledge the counterterms
presented in ref.~\cite{Weinzierl:2003fx} have not yet been integrated over
the corresponding phase spaces.

\section{An application: the $\CF\TR$ part of 
$\epem\to 2$ jets\label{sec:CFTR}}
In this section, we shall apply the subtraction procedure
discussed in sect.~\ref{sec:NNLO} to a physical case, namely the
contribution proportional to the colour factor $\CF\TR$ of the 
dijet cross section in $\epem$ collisions. This is a relatively
simple part of the complete calculation of an observable to NNLO,
but it allows us to discuss the practical implementation of most
of the features of the subtraction procedure we propose in this 
paper. The branching kernels that we shall introduce are
universal, i.e. they can be used in any other computations where
they are relevant. We shall also define precisely the phase spaces
needed to integrate the above kernels over the variables of 
unresolved partons, i.e. the quantities $d\tilde{\psp}$ used in
the formal manipulations of sect.~\ref{sec:NNLO}. We shall show 
that the subtraction procedure leads to the expected KLN
cancellation, and that the numerical integration of the finite
remainder gives a result in excellent agreement with that 
obtained in ref.~\cite{Anastasiou:2004qd}.

Although not necessary for the implementation of eq.~(\ref{jetxsecsub}),
in this section we shall use partial fractioning to deal with the
eikonal factors associated with soft singularities, and combine them
with the corresponding collinear factors by using the {\em same} 
kinematics in the hard matrix elements that factorize. Since such
kinematics will also enter the measurement functions appearing
in the counterterms, all of the manipulations involving the hard matrix
elements will also apply to the measurement functions; for this reason,
the latter will be left implicit in the notation.

We shall reinstate here the notation commonly used 
for QCD amplitudes, which can be written as vectors in a colour space
including the coupling constant. Thus, eq.~(\ref{ampm}) is
now rewritten as
\beq
\colket{\cm_{a_1a_2\dots a_m}}=\colket{\cmt_{a_1a_2\dots a_m}}+
\colket{\cmo_{a_1a_2\dots a_m}}+\colket{\cmd_{a_1a_2\dots a_m}}+\dots
\eeq
where ${a_1a_2\dots a_m}$ are flavour indices, and particles other than
QCD partons are always understood. When not necessary, flavour labels
and colour vector symbols may also be understood.

UV renormalization is performed in the ${\overline {\rm MS}}$ scheme,
just by expressing
the bare coupling $\asu$ in terms of the renormalized coupling $\as(\mu^2)$
at the renormalization scale $\mu$. We use the following expression
\beq
\label{msbarreno}
\asu\,\mu_0^{2\ep}\,S_{\ep} = \as(\mu^2)\,\mu^{2\ep}
\left[ 1 -  \f{\as(\mu^2)}{2\pi} \;\frac{\beta_0}{\ep} +\dots\right]\,,
\eeq
where $\beta_0=\f{11}{6} \CA-\f{2}{3}\nf\TR$, and
\beq
S_\ep=(4\pi)^\ep e^{-\ep\Euler}
\eeq
is the typical phase space factor in $d=4-2\ep$ dimensions
($\Euler=0.5772...$ being the Euler number).

%%%%%%%%%%%%%%%%%%%%%%%%%%%%%%%%%%%%%%%%%%%%%%%%%%%%%%%%%%%%%%%%%%%
\begin{figure}[htb]
  \begin{center}
      \epsfig{figure=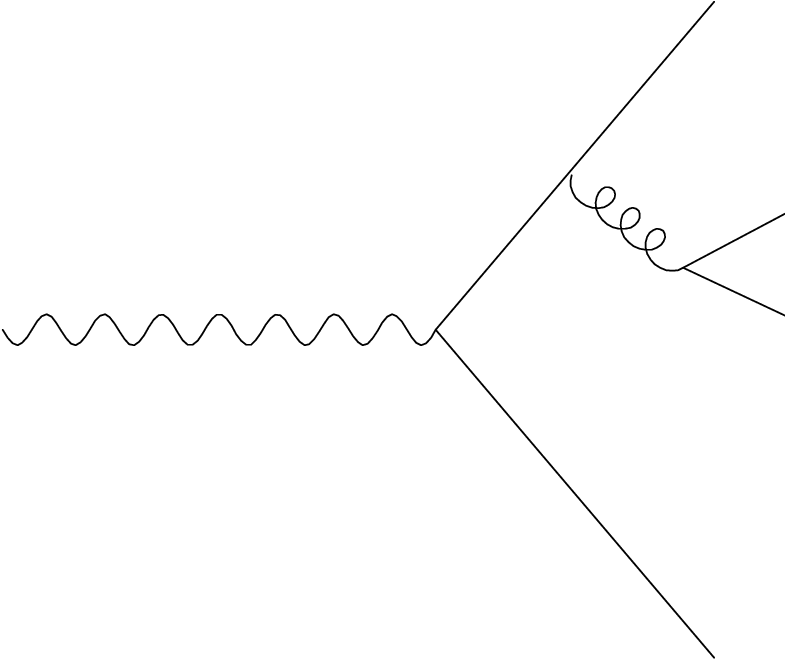,width=0.25\textwidth}
~~~~~~~~~~~~~~~~~~\epsfig{figure=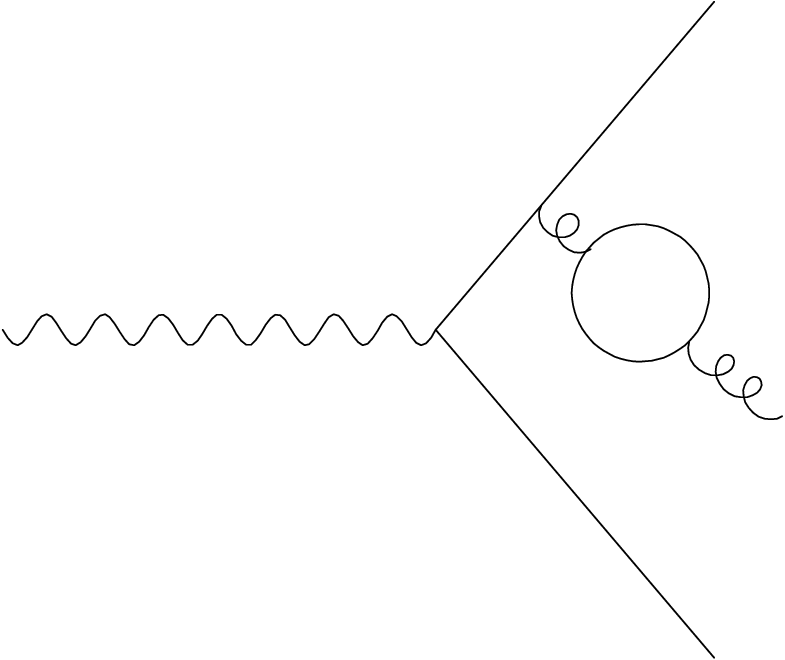,width=0.25\textwidth}
\caption{\label{fig:diagrm} 
Sample diagrams for double-real (left panel) and real-virtual (right panel)
contributions to the $\CF\TR$ part of the dijet cross section.
}
  \end{center}
\end{figure}
%%%%%%%%%%%%%%%%%%%%%%%%%%%%%%%%%%%%%%%%%%%%%%%%%%%%%%%%%%%%%%%%%%%
The diagrams contributing to the $\CF\TR$ part of $\cmt_4$ (i.e., to the
double real) are of the kind of that displayed in the left panel of
fig.~\ref{fig:diagrm}; thus, the singular limits we are interested in are
associated with the branchings $g\to q\bq$ and $q\to q\qp\bqp$ (the
identical-flavour branching $q\to qq\bq$ is also accounted for by considering
$q\to q\qp\bqp$, since its interference contributions are not proportional to
$\CF\TR$). As far as the real-virtual contribution is concerned, a sample
diagram of $\cmo_3$ is depicted in the right panel of fig.~\ref{fig:diagrm}; 
the singular limits of phase-space origin which, according to the discussion 
given in sect.~\ref{sec:NNLOrr}, are obtained upon applying the 
$\E$-prescription, are associated with the branching $q\to qg$.

\subsection{The double-real contribution\label{sec:djrr}}
Thanks to the universality properties of soft and collinear emissions,
any matrix element with
singularities due to the branchings $g\to q\bq$ and $q\to q\qp\bqp$ can 
be used to define the subtraction terms relevant to the double-real 
contribution. Thus, we write the double-real matrix element 
squared as
\beq
rr=\langle\cmt_{\rrtbrc\dots a_{n+2}}
\colket{\cmt_{\rrtbrc\dots a_{n+2}}}\,,
\label{djrr}
\eeq
where the labels imply that the four-momenta of $\qp$, $\bqp$
and $q$ are $k_1$, $k_2$ and $k_3$ respectively. When applying the
$\E$-prescription to eq.~(\ref{djrr}) we obtain from
eqs.~(\ref{tfmodef}) and~(\ref{tfmtdef})
\beqn
\trrmt&=&rr(CC)+rr(SS)-rr(CC\oplus SS)
\nonumber \\*
&-&rr(C\oplus C)-rr(C\oplus SS)+rr(C\oplus C\oplus SS)\,,
\label{djtrrmt}
\\
\trrmo&=&rr(C)\,,
\label{djtrrmo}
\eeqn
where clearly $C\equiv C_{12}$, $CC\equiv CC_{123}$, and so forth.
In order to compute explicitly the quantities that appear in 
eqs.~(\ref{djtrrmt}) and~(\ref{djtrrmo}), we shall use the results
of ref.~\cite{Catani:1999ss}
(see also refs.~\cite{Campbell:1997hg,DelDuca:1999ha}).
The asymptotic behaviours obtained through
successive iterations of $\ENNLO$, characterized by the $\oplus$ symbol,
can be freely defined to a certain extent. We give such definitions
within a given parametrization of the phase space, which we shall
introduce in the next subsection.

\subsubsection{Choices of phase spaces\label{sec:djrrps}}
As discussed in sect.~\ref{sec:NNLOrr} (see in particular 
eq.~(\ref{rrsubt})), the definition of the double-real subtraction
terms implies the necessity of defining $d\tilde{\psp}_{n+2}^{-2}$
and $d\tilde{\psp}_{n+2}^{-1}$ which are related to some extent.
As a preliminary step, we make here the choice of associating 
the same $d\tilde{\psp}_{n+2}^{-2}$ with all of the terms that 
appear in eq.~(\ref{djtrrmt}); this is in principle not necessary
(see eq.~(\ref{unprrmt})), but we find it non restrictive in
the computations that follow (more complicated kernels {\em may}
require a different choice). We then write the exact $(n+2)$-body
phase space using eq.~(\ref{ps:phspfact})
\beq
d\psp_{n+2}=
\frac{ds_{123}}{2\pi}\,
d\psp_{n}(123)\,d\psp_3(1,2,3)\,,
\label{djrrEpst}
\eeq
where we use the momenta labels rather than the four-momenta to shorten
the notation (consistently with eq.~(\ref{psfactFKS}), $123$ means
$k_1+k_2+k_3$), and
\beq
s_{123}=(k_1+k_2+k_3)^2\,.
\eeq
As in the case of NLO computations, eq.~(\ref{djrrEpst}) is unsuited for 
the analytical integration over the variables of unresolved partons since
$123$, which enters the phase space associated with the non-singular part 
of the matrix element, has an off-shellness $s_{123}$, while the reduced
matrix element which corresponds to such non-singular part has all of the
final-state QCD partons with zero mass. We follow the same strategy 
as in eq.~(\ref{psfactFKSs}), i.e. that of 
ref.~\cite{Frixione:1995ms}: we introduce a 
four-momentum $\widetilde{123}$ with invariant mass equal to 
zero\footnote{$\widetilde{123}$ can be defined to have the same
three-momentum as $123$ in the $\epem$ rest frame, and zero mass. 
However, the precise definition is irrelevant in what follows.}, and define
\beq
d\tilde{\psp}_{n+2}^{-2}=
d\psp_{n}(\widetilde{123})\,
\frac{ds_{123}}{2\pi}\,d\psp_3(1,2,3)\,.
\label{psrrmt}
\eeq
This definition is equivalent to that in eq.~(\ref{psfactFKSs}),
and differs from that in eq.~(\ref{psfactCS}) in that no spectator
is used to keep the branching parton on shell.

For the definition of $d\tilde{\psp}_{n+2}^{-1}$, we use 
again eq.~(\ref{ps:phspfact})
\beq
d\psp_{n+2}=
\frac{ds_{12}}{2\pi}\,
d\psp_{n+1}(12)\,d\psp_2(1,2)\,,
\label{djrrEpso}
\eeq
from which the analogue of eq.~(\ref{psrrmt}) can readily follow:
\beq
d\tilde{\psp}_{n+2}^{-1}=
d\psp_{n+1}(\widetilde{12})\,\frac{ds_{12}}{2\pi}\,d\psp_2(1,2)\,.
\label{psrrmow}
\eeq
There is however a subtlety: as discussed in sect.~\ref{sec:NNLOrr},
the choice of $d\tilde{\psp}_{n+2}^{-1}$ must match that of
$d\tilde{\psp}_{n+2}^{-2}$. In order to see how this can happen,
we use eq.~(\ref{ps:threebdso}) to write
\beq
d\tilde{\psp}_{n+2}^{-2}=
d\psp_{n}(\widetilde{123})\,
\frac{ds_{123}}{2\pi}\,\frac{ds_{12}}{2\pi}\,
d\psp_2(3,12)\,d\psp_2(1,2)\,.
\label{tbtemp}
\eeq
The part relevant to the $12\to 1+2$ branching is identical in 
eqs.~(\ref{psrrmow}) and~(\ref{tbtemp}), except for the fact that the 
invariant mass of the $12$ system, $s_{12}$, enters $d\psp_2(3,12)$, and 
not only $d\psp_2(1,2)$, in eq.~(\ref{tbtemp}). If this dependence could 
be safely neglected in the limit $s_{12}\to 0$, i.e. in the singular region 
in which $d\tilde{\psp}_{n+2}^{-1}$ and $d\tilde{\psp}_{n+2}^{-2}$ must 
match, the choice of eq.~(\ref{psrrmow}) would be appropriate. However, 
as can be seen from eqs.~(\ref{ps:tdsoinn}) and~(\ref{ps:gramsa}), 
$d\psp_2(3,12)$ contains the factor
\beq
\left(z_3 z_{12} s_{123} - z_3 s_{12}\right)^{-\ep}\,,
\label{psreg}
\eeq
which acts as a regulator in the integrations in $d\psp_2(3,12)$;
since these are in general divergent, the regulator affects finite 
(and possibly also divergent) terms, and thus cannot be ignored even if the 
limit $s_{12}\to 0$ is considered. The regulator of eq.~(\ref{psreg})
is implicit in $d\psp_{n+1}(12)$ in eq.~(\ref{djrrEpso}); the
effect of neglecting it as done by defining $d\tilde{\psp}_{n+2}^{-1}$
in eq.~(\ref{djrrEpso}) therefore leads to neglecting contributions
to the cross section if $d\psp_{n+1}(\widetilde{12})$ is involved
in the integration of divergent terms. This is what happens, since 
the subtracted real-virtual contribution is indeed divergent.
This problem may seem to be cured by
inserting the regulator of eq.~(\ref{psreg}) into the r.h.s. of
eq.~(\ref{psrrmow}). However, the definition of $d\tilde{\psp}_{n+2}^{-1}$
should not depend on whether the branching $12\to 1+2$ is followed
by the branching $123\to 3+12$, which is what the presence of
$s_{123}$ in eq.~(\ref{psreg}) implies. The most general form
for the regulator to be inserted in eq.~(\ref{psrrmow}) can be
deduced by writing
\beq
d\psp_{n+2}=
\frac{ds_{12}}{2\pi}\,\frac{ds_{3\dots(n+2)}}{2\pi}\,
d\psp_2(12,3\dots(n+2))\,d\psp_2(1,2)\,d\psp_n(3,\dots,(n+2))\,.
\label{djrrpstt}
\eeq
The first phase space on the r.h.s. features the regulator
\beq
\left(z_{12} z_{3\dots(n+2)} Q^2 - z_{12} s_{3\dots(n+2)}
- z_{3\dots(n+2)} s_{12}\right)^{-\ep}\equiv
\left(z_{3\dots(n+2)} s_{12}^\Mx\right)^{-\ep}
\left(1-\frac{s_{12}}{s_{12}^\Mx}\right)^{-\ep}\,,
\label{grampsmo}
\eeq
which can be obtained from eqs.~(\ref{ps:twobd}) and~(\ref{ps:twobdgram}),
and where 
\beq
s_{12}^\Mx=\left(z_{12} z_{3\dots(n+2)} Q^2 - z_{12} s_{3\dots(n+2)}\right)/
z_{3\dots(n+2)}\,,
\label{s12maxdef}
\eeq
with \mbox{$Q^2=(k_1+\dots+k_{n+2})^2$}. Eq.~(\ref{grampsmo}) suggests
to replace eq.~(\ref{psrrmow}) with
\beq
d\tilde{\psp}_{n+2}^{-1}=
d\psp_{n+1}(\widetilde{12})\,\frac{ds_{12}}{2\pi}\,
\left(1-\frac{s_{12}}{s_{12}^\Mx}\right)^{-\ep}
d\psp_2(1,2)\,,
\label{psrrmo}
\eeq
where the quantity $s_{12}^\Mx$ is a constant with respect to the integration
in $s_{12}$.

\subsubsection{Computation of the divergent terms\label{sec:djrrdiv}}
In this section, we integrate $\trrmt$ and $\trrmo$ given in
eqs.~(\ref{djtrrmt}) and~(\ref{djtrrmo}) over the phase spaces 
defined in sect.~\ref{sec:djrrps}. We start from the case
of $\trrmt$, and write
\beq
\trrmt=\colbra{\cmt_{q\dots a_{n+2}}}
\bKth_{\rrtbrc}
\colket{\cmt_{q\dots a_{n+2}}}\,.
\label{djrrmt}
\eeq
The kernel $\bKth_{\rrtbrc}$ is a matrix in the colour space, and 
collects all singular behaviours associated with the branching
$q\to\rrtbrc$ according to the combination given in eq.~(\ref{djtrrmt}).
For its explicit construction we shall use the factorization
formulae given in ref.~\cite{Catani:1999ss}.
In general, we shall denote the branching kernels as follows
\beq
\bK^{(n)}_{a_1\dots a_n}\;\;\;\;\;\;{\rm for}\;\;\;\;\;\;
S(a_1,\dots,a_n)\to a_1+\dots+a_n\,,
\label{kernnot}
\eeq
where $S(a_1,\dots,a_n)$ is the parton flavour which matches the
combination $a_1+\dots+a_n$. In this paper, only $n=2,3$ will be
considered. Notice that the notation of eq.~(\ref{kernnot}) does
not coincide with the one typically used for two-parton branchings,
where the Altarelli-Parisi kernel is denoted by $P_{a_1S(a_1,a_2)}$.
There is a remarkable simplification that occurs in the computation
of eq.~(\ref{djrrmt}); namely, one can prove that (see app.~\ref{sec:simpl})
\beq
rr(SS)=rr(C\oplus SS)\,,\;\;\;\;\;\;\;\;
rr(CC\oplus SS)=rr(C\oplus C\oplus SS)\,.
\eeq
Thus
\beq
\trrmt=rr(CC)-rr(C\oplus C)\,.
\label{djtrrmtsim}
\eeq
Since in the collinear limits colour correlations do not appear,
eq.~(\ref{djtrrmtsim}) implies that $\bKth_{\rrtbrc}$ has a trivial
structure in colour space. Using the same normalization in the
factorization formulae as in ref.~\cite{Catani:1999ss} (see 
eq.~(29) there), we thus rewrite the kernel as follows
\beq
\bKth_{\rrtbrc}=
\f{(8\pi\asu\mu_0^{2\ep})^2}{s_{123}^2}
\left(K_{\rrtbrc}^{CC}-K_{\rrtbrc}^{\CpC}\right)\,,
\label{Kthsim}
\eeq
where the first term can be read from eq.~(57) of ref.~\cite{Catani:1999ss}
\beq
K_{\rrtbrc}^{CC}=
\f{1}{2} \, \CF\TR \,\f{s_{123}}{s_{12}}
\Bigg[- \f{t_{12,3}^2}{s_{12}s_{123}}
+\f{4z_3+(z_1-z_2)^2}{z_1+z_2}+(1-2\ep)
\left(z_1+z_2-\f{s_{12}}{s_{123}}\right)\Bigg]\,,
\label{K3cc}
\eeq
and
\beq
t_{ij,k} \equiv 2 \;\f{z_i s_{jk}-z_j s_{ik}}{z_i+z_j} +
\f{z_i-z_j}{z_i+z_j} \,s_{ij}\,.
\label{tvar}
\eeq
The quantity $K_{\rrtbrc}^{\CpC}$ in eq.~(\ref{Kthsim}) corresponds
to the successive branchings $q\to q_3g$, $g\to\qp_1\bqp_2$. We can 
therefore construct it by using the polarized Altarelli-Parisi kernels
$\Ph_{gq}^{\mu\nu}$ and $\Ph_{q\bq}^{\mu\nu}$, as shown in
app.~\ref{sec:simpl}. The result is 
\beq
K^{\CpC}_{\rrtbrc}=\f{1}{2} \, 
\CF \TR \,\f{s_{123}}{s_{12}}
\Bigg[- \f{t_{12,3}^{(S)^2}}{s_{12}s_{123}}
+\f{4z_3+(z_1-z_2)^2}{z_1+z_2}+(1-2\ep)\left(z_1+z_2\right)\Bigg]\,,
\label{K3cpc}
\eeq
where
\beq
t_{ij,k}^{(S)}\equiv 2 \;\f{z_i s_{jk}-z_j s_{ik}}{z_i+z_j}\,.
\eeq
We note that the result in eq.~(\ref{K3cpc}) corresponds to the naive 
$s_{12}\to 0$ limit of $K_{\rrtbrc}^{CC}$ in eq.~(\ref{K3cc}),
since $t_{12,3}\to t_{12,3}^{(S)}\propto\sqrt{s_{12}}$ in such a limit
(see app.~\ref{sec:phsp}).
The kernel $\bKth_{\rrtbrc}$ is therefore manifestly simpler than
its two contributions in eq.~(\ref{Kthsim}); using the results of
eqs.~(\ref{K3cc}) and~(\ref{K3cpc}), we find
\beq
\bKth_{\rrtbrc}=
\f{(8\pi\asu\mu_0^{2\ep})^2}{s_{123}^2}\,
\f{1}{2}\CF\TR \f{s_{123}}{s_{12}}
\Bigg[- \f{t_{12,3}^2-t_{12,3}^{(S)^2}}{s_{12}s_{123}}
-(1-2\ep)\f{s_{12}}{s_{123}}\Bigg]\,.
\label{Kthfin}
\eeq
We can now obtain the analytical expressions of the $1/\ep$ poles in
$\trrmt$. Using eq.~(\ref{psrrmt})
\beq
\int\trrmt d\tilde{\psp}_{n+2}^{-2}=
\abs{\cmt_{q\dots a_{n+2}}}^2 d\psp_{n}
\int \frac{ds_{123}}{2\pi}\bKth_{\rrtbrc}\,d\psp_3(1,2,3)\,.
\label{djrrmtI}
\eeq
We use the parametrization of eq.~(\ref{ps:pstbso}) to perform the
integration in eq.~(\ref{djrrmtI}). The invariants are expressed in
terms of the integration variables according to eqs.~(\ref{ps:sot})
and~(\ref{ps:stt}); however, one must be careful when replacing
these expressions into $t_{12,3}^{(S)}$, since in doing so
finite terms are generated when $s_{12}\to 0$; such terms should not
appear, since they have been explicitly neglected when working
$K^{\CpC}_{\rrtbrc}$ out. To avoid this, we use
\beq
z_1 s_{23}-z_2 s_{13}\to \sqrt{s_{12}}\xi_0\,,
\;\;\;\;\;\;\;\;
\xi_0=2\sqrt{z_3 \zeta_2(1-\zeta_2)s_{123}(1-z_3)}\, x
\eeq
in the computation of $t_{12,3}^{(S)}$. In a more general case,
the replacements
\beq
s_{13}\to s_{123}(1-\zeta_2)\,,\;\;\;\;\;\;\;\;
s_{23}=s_{123}\, \zeta_2
\eeq
should also be made. The result of the integral in 
eq.~(\ref{djrrmtI}) is
\beq
\int \frac{ds_{123}}{2\pi}\bKth_{\rrtbrc}\,d\psp_3(1,2,3)=
\left(\f{\as}{2\pi}\right)^2\CF\TR 
\left(\f{s_{123}^{\Mx}}{\mu^2}\right)^{-2\ep}
\f{1}{6\ep}\left(1+\f{31}{6}\ep\right)
+{\cal O(\ep)}\,,
\label{intKth}
\eeq
where we denoted by $s_{123}^{\Mx}$ the upper limit of the integration
in $s_{123}$, whose form does not need be specified here; a definite
choice will be made in sect.~\ref{sec:res}.

We now turn to the case of $\trrmo$, and write the analogue of
eq.~(\ref{djrrmt})
\beq
\trrmo=\colbra{\cmt_{gq_3\dots a_{n+2}}}
\bKtw_{\rrobrc}
\colket{\cmt_{gq_3\dots a_{n+2}}}\,.
\label{djrrmo}
\eeq
As in the case of $\bKth_{\rrtbrc}$, the kernel $\bKtw_{\rrobrc}$
is purely collinear (see eq.~(\ref{djtrrmo})), and therefore has a 
trivial colour structure. Using again the factorization formulae
with the normalization of ref.~\cite{Catani:1999ss} (see eq.~(7)
there), we have
\beq
\langle\mu|\bKtw_{\rrobrc}|\nu\rangle
=\f{8\pi\asu\mu_0^{2\ep}}{s_{12}}\Ph_{q\bq}^{\mu\nu}\,,
\label{Ktw}
\eeq
where $\Ph_{q\bq}^{\mu\nu}$ is given in eq.~(\ref{Pqq}). 
Spin correlations are essential for the kernel
$\bKtw$ to be a local subtraction counterterm,
but they do not contribute to the analytic integration which follows,
where the spin-dependent collinear kernel $\Ph_{q\bq}^{\mu\nu}$
can be replaced by
\beq
\langle P_{q\bq}\rangle=\TR\Bigg[1-\frac{2z(1-z)}{1-\ep}\Bigg]\, .
\eeq
Using eq.~(\ref{psrrmo}) we obtain
\beq
\int\trrmo d\tilde{\psp}_{n+2}^{-1}=
\abs{\cmt_{gq_3\dots a_{n+2}}}^2 d\psp_{n+1}
\int \frac{ds_{12}}{2\pi}\bKtw_{\rrobrc}
\left(1-\frac{s_{12}}{s_{12}^\Mx}\right)^{-\ep}d\psp_2(1,2)\,.
\label{djrrmoI}
\eeq
The integral can be performed straightforwardly, with the result
\beqn
&&\int \frac{ds_{12}}{2\pi}\bKtw_{\rrobrc}
\left(1-\frac{s_{12}}{s_{12}^\Mx}\right)^{-\ep}d\psp_2(1,2)=
\f{\as}{2\pi}\, \f{e^{\ep\Euler}}{\Gamma(1-\ep)}\,
\left(\f{\mu^2}{s_{12}^{\Mx}}\right)^{\ep}\TR\left(-\frac{1}{\ep}\right)
\nonumber \\*&&\phantom{aaaaaaaaaaa}\times
\Bigg[\f{2}{3}+\f{10}{9} \ep+\left(\f{56}{27}-\f{\pi^2}{9}\right)\ep^2
+\left(\f{328}{81}-\f{5}{27}\pi^2-\f{4}{3}\zeta_3\right)\ep^3
\Bigg]
\nonumber \\*&&\phantom{aaaaaaaaaaa}\times
\left(1-\f{\pi^2}{6}\ep^2-2\ep^3\zeta_3\right)
+{\cal O}(\ep^3)\,,
\label{intKtw}
\eeqn
where all the terms up to ${\cal O}(\ep^2)$ have been kept, since 
eq.~(\ref{intKtw}) will be used to define the subtracted real-virtual 
contribution as given in eq.~(\ref{rvsubt}), which will generate
further poles up to $1/\ep^2$.

\subsection{The real-virtual contribution\label{sec:djrv}}
In this section, we shall construct the subtracted real-virtual
contribution, defined in eq.~(\ref{rvsubt}). As discussed there,
$rv^{(s)}$ has no explicit $1/\vep$ poles, since those resulting
from the one-loop integrals contributing to the (unsubtracted)
real-virtual contribution $rv$ are cancelled by those of $\trrmo$.
We can show this explicitly for the $\CF\TR$ part of the $\epem$ 
dijet cross section: the contribution of 
\beq
rv=\langle\cmt_{gq\bq}\colket{\cmo_{gq\bq}}+
\langle\cmo_{gq\bq}\colket{\cmt_{gq\bq}}
\label{djrv}
\eeq
to this colour factor is entirely due to the $\TR$ part resulting
from UV renormalization:
\beq
rv|_{\TR}=\f{\as}{2\pi}\f{2}{3}\f{\TR\nf}{\vep}
\abs{\cmt_{gq\bq}}^2.
\label{djrvTR}
\eeq
By using eqs.~(\ref{rvphsp}), (\ref{djrrmoI}), and~(\ref{intKtw}),
with $gq_3\dots a_{n+2}\equiv gq\bq$ and summing over $\nf$ quark
flavours, it is apparent that $rv^{(s)}$ is indeed free of explicit
$1/\vep$ poles:
\beqn
rv^{(s)}d\psp_3&\equiv& rv|_{\TR}d\psp_3+
\sum_{flav}\int\trrmo d\tilde{\psp}_4^{-1}=
\f{\as}{2\pi}\f{\TR\nf}{\vep}\abs{\cmt_{gq\bq}}^2\,d\psp_3
\nonumber\\*&&\times
\Bigg\{\f{2}{3}-\left(\f{\mu^2}{s_{12}^{\Mx}}\right)^{\ep}\,
\f{e^{\ep\Euler}}{\Gamma(1-\ep)}\Bigg[\f{2}{3}+\f{10}{9} \ep+
\left(\f{56}{27}-\f{\pi^2}{9}\right)\ep^2
\nonumber\\*&&\phantom{\times}
+\left(\f{328}{81}-\f{5}{27}\pi^2-\f{4}{3}\zeta_3\right)\ep^3\Bigg]
\left(1-\f{\pi^2}{6}\ep^2-2\ep^3\zeta_3\right)\Bigg\}
+{\cal O}(\ep^3)\phantom{aaaaa}
\label{djrvsub}
\\*&=&
-\f{\as}{2\pi}\f{2\TR\nf}{9}
\left(5+3\log\left(\f{\mu^2}{s_{12}^{\Mx}}\right)\right)
\abs{\cmt_{gq\bq}}^2\,d\psp_3+{\cal O}(\ep)\,.
\label{djrvsubfin}
\eeqn
We underline the presence of $s_{12}^{\Mx}$ in eqs.~(\ref{djrvsub})
and~(\ref{djrvsubfin}), due to the integration performed in 
eqs.~(\ref{djrrmoI}) and~(\ref{intKtw}), and to the choice of phase 
space of eq.~(\ref{psrrmo}). Its specific form depends on the
branchings involved in the construction of $\trrmo$, and can
be read from eq.~(\ref{s12maxdef}); we shall write it explicitly
in the following, after choosing a parametrization for the phase
spaces $d\psp_{n+1}$ and $d\tilde{\psp}_{n+1}$.

Implicit divergences remain in eq.~(\ref{djrvsub}), whose analytic 
computation requires the definition of the subtraction counterterm
$\widetilde{rv}^{(s)}$ which appears in eq.~(\ref{rvfinal}). It is clear
that these divergences are those of the matrix element $\cmt_{gq\bq}$
which factorizes in eq.~(\ref{djrvsub}) and thus are due to the 
branching $q\to qg$. In order to study them with full generality, we consider
\beq
rv^{(0)}=\langle\cmt_{gq\dots a_{n+1}}
\colket{\cmt_{gq\dots a_{n+1}}}\,,
\label{djrvz}
\eeq
from which we construct the subtraction counterterm by applying the
$\E$-prescription:
\beq
\widetilde{rv}^{(0)}=rv^{(0)}(C)+rv^{(0)}(S)-rv^{(0)}(SC).
\label{djrvzsub}
\eeq
Clearly, $\widetilde{rv}^{(s)}$ is obtained by multiplying
$\widetilde{rv}^{(0)}$ by the r.h.s. of eq.~(\ref{djrvsub}),
divided by $\abs{\cmt_{gq\bq}}^2\,d\psp_3$; the same manipulations
can be carried out on the r.h.s. of eq.~(\ref{djrvsubfin}), in
this way obtaining the local subtraction counterterm to be 
used in the numerical computation of the integral that appears
in the first term on the r.h.s. of eq.~(\ref{rvfinal}), or in
the second line on the r.h.s of eq.~(\ref{jetxsecsub}).
We stress that eq.~(\ref{djrvzsub}) coincides with eq.~(\ref{ftilNLO}),
since the subtraction of the phase-space singularities of $rv^{(s)}$
is identical to that performed in the context of an NLO computation.
In what follows, similarly to what done in sect.~\ref{sec:djrr},
we shall choose the same parametrization of the phase space for
the three terms on the r.h.s. of eq.~(\ref{djrvzsub}), whose sum
will therefore be essentially identical to one of the dipole kernels
of ref.~\cite{Catani:1996vz}. As before, we shall however not use the 
dipole parametrization for the phase space, and our integrated kernel
will thus be different from that of ref.~\cite{Catani:1996vz}.
We obtain
\beq
\widetilde{rv}^{(0)}=\colbra{\cmt_{q\dots a_{n+1}}}
\bKtw_{qg}
\colket{\cmt_{q\dots a_{n+1}}}\,,
\label{djrvzfac}
\eeq
where the kernel has now a non-trivial colour structure
\beq
\bKtw_{qg}=-\f{8\pi\mu_0^{2\ep}\asu}{s_{qg}}\sum_{k\ne q}
\bVtw_{qg,k}\,\bT_q\mydot\bT_k\,.
\eeq
Here
\beqn
\bVtw_{qg,k}&=&\f{2p_q\mydot p_k}{(p_q+p_k)\mydot p_g}+
(1-z)(1-\ep),
\label{Vqg}
\\
z&=&\f{p_q\mydot p_k}{(p_q+p_g)\mydot p_k}\,,
\label{rvzdef}
\eeqn
and in order to obtain the result in eq.~(\ref{Vqg}) we have decomposed the 
eikonal factor that appears in $rv^{(0)}(S)$ as shown
in eq.~(\ref{eikid})\footnote{We note that in this calculation
the kernel needed to construct $\widetilde{rv}^{(s)}$
coincides with the NLO one owing to eq.~(\ref{djrvTR}).
In more general cases, a formula similar to (\ref{djrvzsub})
for $\widetilde{rv}^{(s)}$ still holds,
but $rv^{(s)}(C)$,  $rv^{(s)}(S)$ and $rv^{(s)}(SC)$ will have to be computed
using the results of refs.~\cite{bern,Kosower:1999rx,Catani:2000pi}.}.

According to our master subtraction formula, eq.~(\ref{jetxsecsub}), we
have now to integrate $\widetilde{rv}^{(s)}$ over the phase space, and
thus we have to choose a parametrization for $d\tilde{\psp}_{n+1}$.
A possibility is that of using again the form of eq.~(\ref{psrrmo}),
which with the parton labeling of the case at hand would read as follows
\beq
d\tilde{\psp}_{n+1}=
d\psp_n(\widetilde{qg})\,\frac{ds_{qg}}{2\pi}\,
\left(1-\frac{s_{qg}}{s_{qg}^\Mx}\right)^{-\ep}
d\psp_2(q,g)\,.
\label{psrvtil}
\eeq
On the other hand, the present case is simpler than that discussed in
sect.~\ref{sec:djrr}. The integration of $\widetilde{rv}^{(s)}$ over 
$d\tilde{\psp}_{n+1}$ will not be followed by another analytical integration; 
thus, the regulator that appears explicitly in eq.~(\ref{psrvtil}) is
not actually necessary. Furthermore, its presence would require the
presence of an analogous regulator for $s_{123}$ in eq.~(\ref{psrrmt}).
Thus, in order to simplify as much as possible the analytic computations,
we rather adopt the following form
\beq
d\tilde{\psp}_{n+1}=
d\psp_n(\widetilde{qg})\,\frac{ds_{qg}}{2\pi}\,
d\psp_2(q,g)\,.
\label{psrvfin}
\eeq
Having made this choice, we can deduce the expression of $s_{12}^{\Mx}$ 
from eq.~(\ref{s12maxdef}):
\beq
s_{12}^{\Mx}=(1-z)s_{qg}\,.
\label{s12maxch}
\eeq
In order to integrate the kernel $\bKtw_{qg}$ over $ds_{qg}d\psp_2(q,g)$,
we use eq.~(\ref{ps:twobd}), choosing the reference four-vector $n$ there
to coincide with $p_k$; in such a way, the variable $z_1$ of 
eq.~(\ref{ps:zidef}) coincides with $z$ of eq.~(\ref{rvzdef}).
We also define
\beq
Q_{qgk}^2=(p_q+p_g+p_k)^2\,,
\label{Qqgkdef}
\eeq
which is treated as a constant during the integration. We obtain
\beqn
\int \widetilde{rv}^{(0)}d\tilde{\psp}_{n+1}&=&
-\f{\as}{2\pi}\,\f{e^{\ep\Euler}}{\Gamma(1-\ep)}
\sum_{k\ne q}\left(\f{\mu^2}{Q_{qgk}^2}\right)^{\ep}
I_{qg,k}^{(A)}\nn\\
&\times &\colbra{\cmt_{q\dots a_{n+1}}}\bT_q\mydot\bT_k
\colket{\cmt_{q\dots a_{n+1}}}\,d\psp_n\,.
\label{rv0int}
\eeqn
In eq.~(\ref{djrvsub}) we also need
\beqn
\int \widetilde{rv}^{(0)}\left(\f{\mu^2}{s_{12}^\Mx}\right)^{\ep}
d\tilde{\psp}_{n+1}&=&
-\f{\as}{2\pi}\,\f{e^{\ep\Euler}}{\Gamma(1-\ep)}
\sum_{k\ne q}\left(\f{\mu^2}{Q_{qgk}^2}\right)^{\ep}
I_{qg,k}^{(B)}\nn\\
&\times &\colbra{\cmt_{q\dots a_{n+1}}}\bT_q\mydot\bT_k
\colket{\cmt_{q\dots a_{n+1}}}\,d\psp_n\,,
\label{rv0intB}
\eeqn
with $s_{12}^\Mx$ given in eq.~(\ref{s12maxch}), and
\beqn
\f{(4\pi)^\ep}{\Gamma(1-\ep)}
\left(\f{\mu^2}{Q_{qgk}^2}\right)^{\ep}I_{qg,k}^{(A)}&=&
8\pi\mu^{2\ep}\int\frac{ds_{qg}}{s_{qg}}\,d\psp_2(q,g)\,
\bVtw_{qg,k}\,,
\\
\f{(4\pi)^\ep}{\Gamma(1-\ep)}
\left(\f{\mu^2}{Q_{qgk}^2}\right)^{2\ep}I_{qg,k}^{(B)}&=&
8\pi\mu^{2\ep}\int\frac{ds_{qg}}{s_{qg}}\,d\psp_2(q,g)\,
\left(\f{\mu^2}{s_{12}^{\Mx}}\right)^{\ep}\bVtw_{qg,k}\,.
\eeqn
A somewhat lengthy computation returns the following results
\beqn
I_{qg,k}^{(A)}&\equiv & \int_0^1 dz \int_0^{\ymax}dy 
\left[\f{2z(1-y)}{1-z+y z}+(1-z)(1-\ep)\right]
\left(z(1-z)\right)^{-\ep} y^{-1-\ep}
\nn\\*
&=&\f{1}{\ep^2}+\f{3}{2}\f{1}{\ep}+\f{7}{2}-\f{\pi^2}{6}-
\log(\ymax)\left(\log(\ymax)+\f{3}{2}\right)-2\Li_2(1-\ymax)
\nonumber\\*
&+&\ep\Bigg[7-\f{\pi^2}{4}-2\zeta_3+\log(\ymax)
\left(\f{\pi^2}{3}-\f{7}{2}\right)-2\log(\ymax)\Li_2(\ymax)
\nonumber\\*&&
+\log^2 (\ymax)\left(\f{3}{4}-2\log(1-\ymax)+\f{2}{3}\log(\ymax)\right)
\nonumber\\*&&
-2\Li_3(1-\ymax)
+2\Li_3\left(\f{\ymax-1}{\ymax}\right)\Bigg]+{\cal O}(\ep^2)\,,
\label{IAres}
\\
I_{qg,k}^{(B)}&\equiv & \int_0^1 dz \int_0^{\ymax}dy 
\left[\f{2z(1-y)}{1-z+y z}+(1-z)(1-\ep)\right]
z^{-\ep}(1-z)^{-2\ep} y^{-1-2\ep}\nn
\\*
&=&\f{1}{4\ep^2}+\f{3}{4\ep}+\f{21}{8}-
\log(\ymax)\left(\f{3}{2}+\log(\ymax)\right)
-2\Li_2(1-\ymax)
\nonumber\\*
&+&\ep\Bigg[\f{127}{16}-\f{\pi^2}{4}-\zeta_3
+\log(\ymax)\left(\f{\pi^2}{3}-\f{21}{4}\right)
-6\log(\ymax)\Li_2(\ymax)
\nonumber\\*&&
+\log^2(\ymax)\left(\f{3}{2}-4\log(1-\ymax)+\f{4}{3}\log(\ymax)\right)
\nonumber\\*&&
-2\Li_3(1-\ymax)+4\Li_3\left(\f{\ymax-1}{\ymax}\right)
+4\Li_3(\ymax)\Bigg]+{\cal O}(\ep^2)\,,
\label{IBres}
\eeqn
where we have introduced the {\em arbitrary} parameter $\ymax$, which
defines the upper limit of the $s_{qg}$ integration
\beq
s_{qg}\le\ymax\,Q_{qgk}^2\,,\;\;\;\;\;\;\;\;
0<\ymax\le 1\,.
\label{sqgupp}
\eeq
The physical results must not depend on $\ymax$, whose variation amounts
to changing finite contributions to the subtraction counterterms; whether
this condition is fulfilled represents a powerful check on the correctness
of the subtraction procedure. The parameter $\ymax$ is the analogue of the
free parameters $\xi_{cut}$, $\delta_{\sss I}$ and $\delta_{\sss O}$ 
of ref.~\cite{Frixione:1995ms}.

\subsection{Results\label{sec:res}}
In this section, we shall use the results obtained in sects.~\ref{sec:djrr}
and~\ref{sec:djrv} in order to implement our master subtraction formula
eq.~(\ref{jetxsecsub}). As a preliminary step, we need the $\CF\TR$ part
of the double-virtual contribution
\beq
vv=\langle\cmt_{q\bq}\colket{\cmd_{q\bq}}+
\langle\cmd_{q\bq}\colket{\cmt_{q\bq}}\,,
\label{djvv}
\eeq
which can be straightforwardly obtained from ref.~\cite{Matsuura:1988sm}:
\beqn
vv|_{\CF\TR}&=&\CF\TR\nf
\left(\f{\as}{2\pi}\right)^2 \abs{\cmt_{q\bq}}^2
\nonumber\\*
&\times&\Bigg\{
\left(\f{Q^2}{\mu^2}\right)^{-2\ep}\Bigg[\f{1}{3\ep^3}+\f{14}{9\ep^2}+
\left(\f{353}{54}-\f{11}{18}\pi^2\right)\f{1}{\ep}+\f{7541}{324}
-\f{77}{27}\pi^2-\f{26}{9}\zeta_3\Bigg]
\nonumber\\*&&
+\left(\f{Q^2}{\mu^2}\right)^{-\ep}\Bigg[-\f{4}{3\ep^3}-\f{2}{\ep^2}+
\left(\f{7}{9}\pi^2-\f{16}{3}\right)\f{1}{\ep}-\f{32}{3}+\f{7}{6}\pi^2+
\f{28}{9}\zeta_3\Bigg]
\Bigg\}\,,\phantom{aaaaaa}
\eeqn
where $Q^2$ is the $\epem$ c.m. energy squared.
The subtraction counterterms in the first and second integrals on the r.h.s. 
of eq.~(\ref{jetxsecsub}) are constructed using the kernels $\bKtw$ and
$\bKth$ defined in the previous sections. It must be stressed that 
we shall have to subtract $\bKth$ twice, since we have two configurations
contributing to topology $\Tt$: the first is that depicted on the left panel
of fig.~\ref{fig:diagrm}, the second being identical except for the
fact that the gluon is attached to the other quark leg emerging from the
$\gamma\to q\bq$ branching. This also implies that the double-real
contribution to the last line on the r.h.s. of eq.~(\ref{jetxsecsub}) 
will be obtained upon multiplying by two the result presented in 
eq.~(\ref{intKth}) (an overall factor of $\nf$ appears when summing
over flavours). Finally, the largest kinematically-allowed value of 
the virtuality of the branching parton in a $1\to 3$ splitting is $Q^2$. 
For consistency with what done in eq.~(\ref{sqgupp}), we thus set
\beq
s_{123}^{\Mx}=\ymax\,Q^2
\label{s123upp}
\eeq
as the upper limit of the integration over $s_{123}$ performed
in sect.~\ref{sec:djrrdiv}. The real-virtual contribution
$\int\widetilde{rv}^{(s)}d\tilde{\psp}_{n+1}$ can be obtained from 
eqs.~(\ref{djrvsub}), (\ref{rv0int}) and (\ref{rv0intB});
for the dijet cross section,
$q\dots a_{n+1}\equiv q\bq$, and the colour algebra is trivial:
\beq
\colbra{\cmt_{q\bq}}\bT_q\mydot\bT_{\bq}
\colket{\cmt_{q\bq}}=-\CF\abs{\cmt_{q\bq}}^2\,.
\eeq
In the computation of the real-virtual contribution, we also 
necessarily have $Q_{qgk}^2\equiv Q^2$ (see eq.~(\ref{Qqgkdef})).
When putting this all together, we obtain what follows for the
analytic-computed part of the subtraction formula of
eq.~(\ref{jetxsecsub}):
\beqn
&&\int\trrmt d\tilde{\psp}_4^{-2}
+\int\widetilde{rv}^{(s)}d\tilde{\psp}_3
+\int vv d\psp_2\,\Bigg|_{\CF\TR}=
\CF\TR\nf
\left(\f{\as}{2\pi}\right)^2 \abs{\cmt_{q\bq}}^2 d\psp_2
\nonumber\\*&&\phantom{aaaa}
\times\Bigg\{\f{19}{9}-\f{29}{27}\pi^2
-\f{2}{3}\log\ymax
+\log\ymax \left(\f{17}{3}+\f{16}{3}\Li_2(\ymax)\right)
\nonumber\\*&&\phantom{aaaa\times}
+\log^2\ymax\left(\f{11}{9}+\f{8}{3}\log(1-\ymax)-\f{8}{9}\log\ymax\right)
\nonumber\\*&&\phantom{aaaa\times}
+\f{40}{9}\Li_2(1-\ymax)
-\f{8}{3}\Li_3\left(\f{\ymax-1}{\ymax}\right)-\f{16}{3}\Li_3(\ymax)
\nonumber\\*&&\phantom{aaaa\times}
+\left(\f{2}{3}-\f{4}{9}\pi^2+\f{4}{3}\log^2\ymax+2\log\ymax
+\f{8}{3}\Li_2(1-\ymax)\right)\log\f{\mu^2}{Q^2}
\Bigg\}\,.\phantom{aaaaaa}
\label{analytic}
\eeqn
This result is non-divergent, which proves the successful cancellation,
as dictated by the KLN theorem, of the soft and collinear divergences 
which arose in the intermediate steps of the computation. 

The cancellation of the divergences does not guarantee that the finite
parts resulting from the implementation of the subtraction procedure are
all included correctly. In order to check this, we started by computing 
the $\CF\TR$ part of the NNLO contribution to the total hadronic cross 
section, and compared it with the well-known analytic result, which can 
be read from~\cite{sigmatot}
\beqn
R&=&\sum_q e^2_q\Bigg\{1+\left(\f{\as}{2\pi}\right)\f{3}{2}\CF
+\left(\f{\as}{2\pi}\right)^2
\Bigg[-\f{3}{8}\CF^2+\CF\CA\left(\f{123}{8}-11\zeta_3\right)
\nn\\*&&
\phantom{\sum_q e^2_q\Bigg\{1+aaaaa}
+\CF\TR\nf\left(-\f{11}{2}+4\zeta_3\right)\Bigg]+{\cal O}(\as^3)\Bigg\}\,,
\eeqn
where
\beq
R=\f{\sigma(e^+e^-\to {\rm hadrons})}{\sigma(e^+e^-\to \mu^+\mu^-)}\,.
\eeq
Although the physical result must not depend on $\ymax$, the three lines
on the r.h.s. of eq.~(\ref{jetxsecsub}) (which we call 4-parton,
3-parton, and analytic contributions respectively) separately do.
We thus compute the rate for different choices of $\ymax$. The
results are presented in table~\ref{tab:rates}.
%%%%%%%%%%%%%%%%%%%%%%%%%%%%%%%%%%%%%%%%%%%%%%%%%%%%%%%%%%%%%%%%%%%
\begin{table}
\begin{center}
\begin{tabular}{crrrrc}
\hline
$\ymax$ & 4-parton & 3-parton & analytic & total & pull\\
\hline
0.2 & $-$0.692 $\pm$ 0.002 & 1.4241 $\pm$ 0.0004 &  $-$1.4236 
& $-$0.692 $\pm$ 0.002& 0.11\\
0.4 & $-$0.229 $\pm$ 0.004 & 9.8001 $\pm$ 0.0003 & $-$10.2610 
& $-$0.690 $\pm$ 0.004& 0.44\\
0.6 &    0.074 $\pm$ 0.02~$\,$ & 12.3440 $\pm$ 0.0003& $-$13.0753 
& $-$0.657 $\pm$ 0.02~$\,$  & 1.74\\
0.8 &    0.235 $\pm$ 0.005 & 13.3759 $\pm$ 0.0003& $-$14.2989 
& $-$0.688 $\pm$ 0.005& 0.75\\
1.0 &    0.383 $\pm$ 0.004 & 13.8284 $\pm$ 0.0003& $-$14.9005 
& $-$0.689 $\pm$ 0.004& 0.69\\
\hline
\end{tabular}
\caption{\label{tab:rates}
Numerical results for the $\CF\TR\nf$ term of the $R$ ratio. 
The three unphysical contributions, called 4-parton, 3-parton, 
and analytic, correspond to the three lines on the r.h.s. of
eq.~(\ref{jetxsecsub}) respectively; their sums are reported under 
the column ``total''. The pulls are defined in eq.~(\ref{pulldef}).
}
\end{center}
\end{table}
%%%%%%%%%%%%%%%%%%%%%%%%%%%%%%%%%%%%%%%%%%%%%%%%%%%%%%%%%%%%%%%%%%%
As can be seen there, the dependence on $\ymax$ of the three 
contributions to our subtraction formula is fairly large. However,
such a dependence cancels in the sum, within the statistical 
accuracy of the numerical computation. We defined
\beqn
{\rm pull}&=&\f{\abs{{\rm total}-{\rm exact}}}{{\rm error}}\,,
\label{pulldef}
\\
{\rm exact}&=&-\f{11}{2}+4\zeta_3=-0.69177\,,
\eeqn
the error being that due to the numerical computation, reported
under the column ``total'' in table~\ref{tab:rates}. We note that
the 4-parton result changes sign when $\ymax$ is increased,
which implies that the integrand gives both positive and negative 
contributions. As is well known, when such contributions are 
close in absolute value, as when choosing $\ymax=0.6$, the computation
of the integral is affected by a relatively large error; in fact, the 
worst agreement with the exact result is obtained with $\ymax=0.6$. 

We finally computed a proper dijet total rate, by reconstructing
the jets using the JADE algorithm with $y_{cut}=0.1$ (for a discussion on
$\epem$ jet algorithms, see e.g. ref.~\cite{Bethke:1991wk}). Using the
same normalization as in ref.~\cite{Anastasiou:2004qd}, we find for
the NNLO contribution proportional to $\nf$
\beq
1.7998\pm 0.0016\;\;\;\;\;\;\;\;
{\rm with}\;\;\;\;\;\;\;\;
\ymax=0.6\,,
\label{JADEres}
\eeq
which is in nice agreement with the result of ref.~\cite{Anastasiou:2004qd}.
We point out that the small error in eq.~(\ref{JADEres}),
resulting from a run with lower statistics with respect to the results
in table \ref{tab:rates}, is obtained with the value
of $\ymax$ that gives the worst convergence performance in the case 
of the total rate. We verified that, as for total rates, dijet 
cross sections are independent of $\ymax$ within the error
of the numerical computation (for example, with $\ymax=0.4$ we
obtain $1.7992\pm 0.0015$).

\section{Comments\label{sec:comm}}
We have shown in sect.~\ref{sec:CFTR} that our master subtraction
formula, eq.~(\ref{jetxsecsub}), allows us to cancel analytically
the soft and collinear divergences relevant to an NNLO computation.
By using the $\E$-prescription, the subtraction counterterms
$\trrmt$, $\trrmo$, and $\widetilde{rv}^{(s)}$ are constructed
from the basic kernels that account for the matrix element
singularities. The successful implementation of the subtraction
procedure also requires a sensible definition of the phase spaces
used to integrate the counterterms. For convenience, we collect 
here the phase space parametrizations we used, relabeling the
partons where necessary, in such a way that those involved in
the singular branchings have always the smallest labels.
For the $\trrmt$ counterterm to the double-real contribution
we used eq.~(\ref{psrrmt})
\beq
d\tilde{\psp}_{n+2}^{-2}=
d\psp_{n}(\widetilde{123})\,
\frac{ds_{123}}{2\pi}\,d\psp_3(1,2,3)\,,
\label{psrrmt2}
\eeq
relevant to topology $\Tt$ (topology $\Topo$ was not relevant to the case
studied in this paper); we could also use eq.~(\ref{tbtemp}), which is fully 
equivalent and particularly suited to integrate kernels with a structure 
analogous to that of strongly-ordered limits. For the $\trrmo$ counterterm 
we used eq.~(\ref{psrrmo})
\beqn
d\tilde{\psp}_{n+2}^{-1}&=&
d\psp_{n+1}(\widetilde{12})\,\frac{ds_{12}}{2\pi}\,
\left(1-\frac{s_{12}}{s_{12}^\Mx}\right)^{-\ep}
d\psp_2(1,2)
\label{psrrmo2}
\\*
&\equiv&d\psp_{n+1}d\psp_2^{-1}\,.
\label{psptemp}
\eeqn
Equations~(\ref{psrrmo2}) and~(\ref{psptemp}) give an explicit 
expression for $d\psp_2^{-1}$, which needed not be specified 
in eq.~(\ref{rvsubt}). Finally, for the counterterm to the 
subtracted real-virtual contribution we used eq.~(\ref{psrvfin})
\beq
d\tilde{\psp}_{n+1}=
d\psp_n(\widetilde{12})\,\frac{ds_{12}}{2\pi}\,
d\psp_2(1,2)\,.
\label{psrvfin2}
\eeq
The factor that depends on ${s_{12}^\Mx}$ in eq.~(\ref{psrrmo2}) is
crucial for the correct definition of the subtracted real-virtual
contribution. The quantity ${s_{12}^\Mx}$ 
depends on the kinematics of the hard system whose phase space is 
$d\psp_{n+1}(\widetilde{12})$. Such a dependence is in general
non trivial: the difference in the results of eqs.~(\ref{IAres})
and~(\ref{IBres}) is due to it (notice that all of the $1/\vep$ poles
are affected; thus, the failure to include the ${s_{12}^\Mx}$-dependent
regulator in eq.~(\ref{psrrmo2}) would prevent the cancellation 
of singularities). The dependence of ${s_{12}^\Mx}$ on 
the kinematics of the hard system which factorizes needs not be known 
analytically for the computation of $\trrmo d\tilde{\psp}_{n+2}^{-1}$
and for the definition of $rv^{(s)}$. It must be known when 
defining $\widetilde{rv}^{(s)}$, since this quantity has to be
integrated analytically. However, in such a case the dependence
is expected to be relatively simple (see eq.~(\ref{s12maxch})).
This is so because the subtraction term $\widetilde{rv}^{(s)}$
always corresponds to the kinematics of a strongly-ordered limit,
in which the branching of partons $1$ and $2$ will be followed by 
another two-parton branching (in the case studied in sect.~\ref{sec:CFTR},
we had $q\to gq_3$ and $g\to\qp_1\bqp_2$; the case of topology
$\Topo$ will clearly be even simpler).

The phase spaces in eqs.~(\ref{psrrmo2}) and~(\ref{psrvfin2}) will
serve to integrate the $\bKtw$ kernels. This implies that, for a 
given kernel, two different integrals will have to be computed,
because of the presence of the ${s_{12}^\Mx}$-dependent regulator 
in $d\tilde{\psp}_{n+2}^{-1}$. This can clearly be avoided by
inserting the regulator also in eq.~(\ref{psrvfin2}). It should
however be stressed that, in doing so, we are forced to insert an 
analogous regulator (${s_{123}^\Mx}$-dependent) in eq.~(\ref{psrrmt2}).
Thus, the advantage of having to compute half of the integrals relevant
to the $\bKtw$ kernels may be lost because of the additional complications
in the computations of the integrals of the $\bKth$ kernels.

\section{Conclusions\label{sec:concl}}
In this paper, we have proposed a framework for the implementation of a
subtraction formalism that allows us to cancel the soft and collinear
singularities which arise in the intermediate steps of a perturbative
computation at the next-to-next-to-leading order in QCD. The strategy is
analogous to that adopted at the next-to-leading order, which is based on the
definition of subtraction kernels whose form is both observable- and
process-independent. We have introduced a systematic way (the
$\E$-prescription) for defining the kernels, that results naturally in a
two-step procedure.

The first step is the definition of the subtraction counterterms
for the double-real contribution. They are of two different types,
denoted by $\trrmt$ and $\trrmo$; roughly speaking, the former collects
all the pure NNLO-type singularities, whereas the latter is singular
in those soft, collinear, and soft-collinear configurations that are
also relevant to NLO computations. The $\E$-prescription guarantees
the absence of double counting.
In the second step, the term $\trrmo$ is summed to the real-virtual
contribution; this sum is free of explicit $1/\vep$ poles, but is
still divergent, and a further subtraction needs be performed. This
subtraction is completely analogous to that relevant to NLO computations.

Our master formula, eq.~(\ref{jetxsecsub}), thus achieves the 
cancellation of the soft and collinear singularities essentially
by two successive NLO-type subtractions; the first defines the
subtracted real-virtual contribution, and the second removes its
remaining singularities. The singularities that cannot be possibly
obtained in this way are all contained in the subtraction counterterms
$\trrmt$ (the singularities of the double-virtual contribution are
of no concern for the subtraction procedure -- we assume the
relevant matrix elements to be available).

Equation~(\ref{jetxsecsub}) is, as it stands, not sufficient for an actual
numerical computation, since suitable choices of the phase spaces involved are
necessary. We tested our subtraction formalism and the corresponding phase
space choices in the context of a simple application, the calculation of the
contribution proportional to $\CF\TR$ of the dijet cross section in $\epem$
collisions. Although this is clearly a simple example, it is, to the best of
our knowledge, the first application in which NNLO process-independent
subtraction counterterms have been constructed and integrated over the
corresponding phase spaces to achieve an explicit cancellation of soft
and collinear singularities.

We point out that, although the phase-space parametrizations adopted in this
paper are seen to induce reasonably simple analytic integrations, this 
is in part due to the relative simplicity of the kernels needed in
the computation of the $\CF\TR$ part of the $\epem$ dijet cross section.
More complicated kernels may require different parametrizations, 
which we did not consider in the present paper. In particular, the two
NLO-type successive subtractions are liable to be simplified with respect 
to what done here. In any case, the general subtraction formula of
eq.~(\ref{jetxsecsub}) will retain its validity independently of
the specific phase-space parametrizations adopted.

There are a few difficulties that we did not address directly in this
paper; they are not difficulties of principle, but may pose technical 
problems. One interesting feature will emerge when computing the 
$\CF\TR$ part of the three-jet cross section in $\epem$ collisions,
namely the interplay between the $\Tt$ and $\Topo$ topologies which
need be treated separately in the definition of the subtraction
kernels. The case of collisions with one or two initial-state hadrons
will also require a more involved notation (although the subtraction
procedure will be basically unchanged). In general, it is clear that
there is a fair amount of work to be done before the subtraction scheme
proposed here is of any phenomenological use. This implies not only
the definition, and integration over the phase spaces, of the universal
subtraction kernels for all of the partonic branchings possibly occurring 
at NNLO, but also the computation of more two-loop amplitudes,
which is a necessary condition for process-independent formalisms
to be more convenient than observable-specific results.

\section*{Acknowledgements}
We wish to thank Stefano Catani, Lorenzo Magnea, Fabio Maltoni, and 
Michelangelo Mangano for comments on the manuscript, and CERN Theory Division 
for hospitality at various stages of this work. S.F. is grateful to Zoltan 
Kunszt and Adrian Signer for the many discussions they had on this matter 
a while ago, and M.G. wishes to thank Stefano Catani for having introduced 
him to the subject.

\newpage
\appendix
\section{The three-parton kernel $\bKth_{\rrtbrc}$\label{sec:simpl}}
In this appendix we give more details on the construction of the
three-parton kernel $\bKth_{\rrtbrc}$. According to eq.~(\ref{djtrrmt}), 
the singularities due to the branching $q\to q\qp\bqp$ are described at 
NNLO by six terms. 

We start by showing that, of these six terms, only $rr(CC)$ and $r(\CpC)$ 
survive. Let us consider the limit in which the $q\bq$ pair becomes soft. 
In this limit the matrix element squared behaves as~\cite{Catani:1999ss}
\beqn
\label{fss}
rr(SS)&=& (4\pi\asu\mu_0^{2\ep})^2\,
\f{\TR}{(k_1\mydot k_2)^2}
\sum_{i,j=3}^n\la {\cal M}_{a_3...a_n}|{\bom T}_i\cdot {\bom T}_j|
{\cal M}_{a_3...a_n}\ra
\nn\\*
&\times&\f{k_i^\mu}{k_i\mydot (k_1+k_2)}
(-k_1\mydot k_2 g_{\mu\nu}+k_{1\mu}k_{2\nu}+k_{1\nu}k_{2\mu})
\f{k_j^\nu}{k_j\mydot (k_1+k_2)}\,.
\eeqn
In the limit in which $q_1$ and $\bq_2$ become collinear the singular behaviour
is given by
\beqn
\label{fc}
rr(C) &=& (4\pi\asu\mu_0^{2\ep})
\f{1}{k_1\mydot k_2}{\hat P}_{q\bq\mu\nu}(z,\kt)\,
{\cal T}^{\mu\nu}_{ga_3...a_n}
\nn\\
&=& (4\pi\asu\mu_0^{2\ep})\f{\TR}{k_1\mydot k_2}
\left(-g_{\mu\nu}+4z(1-z)\f{\ktmu\ktnu}{\kt^2}\right)
{\cal T}^{\mu\nu}_{ga_3...a_n}\, ,
\eeqn
where ${\cal T}_{ga_3...a_n}$ is the spin polarization tensor obtained 
by squaring the amplitude ${\cal M}_{ga_3...a_n}$ and summing over colours 
and spins except the spin the gluon.

Let us now take the double-soft limit of eq.~(\ref{fc}).
By using soft-gluon insertion rules we find
\beqn
\label{fcpss}
rr(C\oplus SS) &=& -(4\pi\asu\mu_0^{2\ep})^2\,
\f{\TR}{k_1\mydot k_2}
\sum_{i,j=3}^n\la {\cal M}_{a_3...a_n}|{\bom T}_i\cdot {\bom T}_j|
{\cal M}_{a_3...a_n}\ra
\nn\\*
& &~~~~\times\f{k_i^\mu}{k_i\mydot (k_1+k_2)}
\left(-g_{\mu\nu}+4z(1-z)\f{\ktmu\ktnu}{\kt^2}\right)
\f{k_j^\nu}{k_j\mydot (k_1+k_2)}
\nn\\
&=&-(4\pi\asu\mu_0^{2\ep})^2\,
\f{\TR}{(k_1\mydot k_2)^2}
\sum_{i,j=3}^n\la {\cal M}_{a_3...a_n}|{\bom T}_i\cdot {\bom T}_j|
{\cal M}_{a_3...a_n}\ra
\nn\\*
& &~~~~\times\f{k_i^\mu}{k_i\mydot (k_1+k_2)}
\left(-k_1\mydot k_2\,g_{\mu\nu}-2\ktmu\ktnu\right)
\f{k_j^\nu}{k_i\mydot (k_1+k_2)}\, .
\eeqn
where we have used $\kt^2=-2z(1-z)k_1\mydot k_2$.

We now want to show that eq.~(\ref{fcpss}) is equivalent to eq.~(\ref{fss}).
We recall that the transverse vector $\kt$ is defined through the Sudakov 
parametrization 
\beqn
k_1^\mu &=& z\, p^\mu+\kt^\mu-\f{\kt^2}{2z\,p\mydot n}n^\mu\,,
\nn\\
k_2^\mu &=& (1-z)\, p^\mu-\kt^\mu-\f{\kt^2}{2(1-z)p\mydot n}n^\mu\,,
\eeqn
where $p$ is the collinear direction and $n$ is an arbitrary additional 
vector used to define the collinear limit. Performing the replacement
\begin{equation}
\label{repl}
\kt\to zk_1-(1-z)k_2
\end{equation}
we neglect terms proportional to
the collinear direction, that vanish by gauge invariance when contracted
with the on shell matrix element.
Gauge invariance, or equivalently, the fact that the
eikonal current is conserved, can be further
exploited to freely exchange $k_{1\mu}\lra -k_{2\mu}$
in eq.~(\ref{repl}); we can thus perform the replacements\footnote{Note 
that the argument works in the same way starting from
$\kt\to (1-z)k_1-zk_2$.}
\begin{equation}
\ktmu\to k_{1\mu}~~~~{\kt}_\nu\to -k_{2\nu}~~~~{\rm or}~~~
2\ktmu\ktnu\to -k_{1\mu}k_{2\nu}-k_{1\nu}k_{2\mu}
\end{equation}
thus showing the equivalence of eqs.~(\ref{fss}) and (\ref{fcpss}).
We conclude that $rr(C\oplus SS)=rr(SS)$ and
$rr(C\oplus C\oplus SS)=rr(CC\oplus SS)$.

We now turn to the computation of $K_{\rrtbrc}^{\CpC}$, which appears
in eq.~(\ref{Kthsim}). As mentioned there, this quantity corresponds to
the branchings $q\to q_3g$, $g\to\qp_1\bqp_2$, which are described 
by the polarized Altarelli-Parisi kernels
\beqn
\Ph_{gq}^{\mu\nu}(z,\kt)&=&\f{\CF}{2\TR}\, z\, \Ph_{q\bq}^{\mu\nu}
\left(\f{1}{z},\kt\right)\,,
\label{Pgq}
\\
\Ph_{q\bq}^{\mu\nu}(z,\kt)&=&\TR\left[-g^{\mu\nu}+4z(1-z)
\f{\kt^\mu \kt^\nu}{\kt^2}\right]\, .
\label{Pqq}
\eeqn
By using the gluon polarization tensor
\beq
d^{(n)}_{\mu\nu}(k)=-g_{\mu\nu}+\f{k_\mu n_\nu+k_\nu n_\mu}{k\mydot n}
\eeq
we can compute
\beq
K^{\CpC}_{\rrtbrc}=
\f{s_{123}}{s_{12}}\, \Ph_{gq}^{\alpha\beta}(z_1+z_2,\qt)\,
d^{(n)}_{\alpha\mu}(k_1+k_2)\,
\Ph_{q\bq}^{\mu\nu}\left(\f{z_1}{z_1+z_2},\kt\right)
d^{(n)}_{\nu\beta}(k_1+k_2)\,.
\label{CpCdef}
\eeq
The vectors $\qt$ and $\kt$ in eq.~(\ref{CpCdef}) are the transverse vectors
involved in the $q\to q_3g$ and $g\to\qp_1\bqp_2$ collinear splitting
respectively. By using the Sudakov parametrization
\beq
k_i^\mu=z_i p^\mu+\kti^\mu-\f{\kti^2}{z_i}\f{n^\mu}{2p\mydot n}\,,
\;\;\;\;\;\;\;\;i=1,2,3\, ,
\eeq
$p$ being the collinear direction, $\qt$ and $\kt$ can be identified with
\beq
\qt=\ktt\,,\;\;\;\;\;\;
\kt=z_2 \kto-z_1 \ktd \,.
\eeq
We have
\beqn
&&\kt^2=-z_1z_2 s_{12}\,,\;\;\;\;\;\;\;\;
\qt^2=-z_3(z_1+z_2)\left(s_{13}+s_{23}\right)\,,
\\&&
2\qt\mydot\kt=(z_1+z_2)\left(z_2s_{13}-z_1s_{23}\right)\,,
\eeqn
where terms proportional to $s_{12}$ in $\qt^2$ and $2\qt\mydot\kt$ have 
been neglected. Notice that
\beq
d^{(n)}_{\mu\nu}(k_1+k_2)=
-g_{\mu\nu}+\f{p_\mu n_\nu+p_\nu n_\mu}{p\mydot n}+{\cal O}(\kti)
\eeq
and the ${\cal O}(\kti)$ terms can be neglected in the evaluation of 
$K^{\CpC}_{\rrtbrc}$. Performing the algebra we obtain the result
presented in eq.~(\ref{K3cpc}).

\section{Phase spaces\label{sec:phsp}}
This section collects the formulae relevant to the phase spaces
used elsewhere in this paper. We refrain from giving the details
of the derivations, which can all be worked out by using the
techniques of ref.~\cite{Byckling}.
The phase space for the final-state particles of the process
\beq
Q\;\longrightarrow\;k_1+\cdots +k_n
\label{ps:process}
\eeq
is denoted by
\beq
d\psp_n(Q;k_1,\ldots,k_n)=
(2\pi)^d\delta^{(d)}\left(Q-\sum_{i=1}^n k_i\right)
\prod_{i=1}^n \frac{d^{d-1}k_i}{(2\pi)^{d-1}2k_i^0}\,,
\label{ps:phspdef}
\eeq
where $d$ is the number of space-time dimensions; with abuse of notation,
we may not list explicitly on the l.h.s. those momenta not involved in
the necessary manipulations. The $(d-1)$-dimensional Euclidean measure is
\beq
d^{d-1}k_i=\abs{k_i}^{d-2}d\abs{k_i}d\Omega_i^{(d-1)}\,.
\label{ps:Emeas}
\eeq
The angular measure is
\beqn
d\Omega^{(d)}&=&
\Omega^{(d-1)}\left(\sin\theta_{d-1}\right)^{d-2}d\theta_{d-1}
\label{ps:dOmegaone}
\\&=&
\Omega^{(d-2)}\left(\sin\theta_{d-1}\right)^{d-2}d\theta_{d-1}
\left(\sin\theta_{d-2}\right)^{d-3}d\theta_{d-2}
\label{ps:dOmegatwo}
\\&=&
\Omega^{(d-k)}\prod_{i=1}^k
\left(\sin\theta_{d-i}\right)^{d-i-1}d\theta_{d-i}\,,
\label{ps:dOmegak}
\eeqn
with 
\beq
0\le\theta_1\le 2\pi\,,\;\;\;\;\;\;
0\le\theta_i\le\pi\,,\;\;i=2,\ldots,d-1\,,
\eeq
and
\beq
\Omega^{(d)}\equiv \int d\Omega^{(d)} =
\frac{2\pi^{d/2}}{\Gamma(d/2)}\,.
\label{ps:totOmegad}
\eeq
The recursive definitions of eqs.~(\ref{ps:dOmegaone})--(\ref{ps:dOmegak}) 
can be continued analytically to complex values of $d$.
Equation~(\ref{ps:process}) can be rewritten as follows
\beq
\begin{array}{rcl}
Q & \longrightarrow & q+k_{m+1}+\cdots +k_n \\
  &  & \!\bentarrow {k_1+\cdots +k_m}\,,
\end{array}
\label{ps:procdec}
\eeq
where we have introduced the $d$-vector $q$
\beq
q=k_1+\cdots +k_m\,.
\eeq
Equation~(\ref{ps:procdec}) suggests to decompose the $n$-body phase 
space in the following way
\beq
d\psp_n(Q;k_1,\ldots,k_n)=
\frac{dq^2}{2\pi}\,
d\psp_{n-m+1}(Q;q,k_{m+1},\ldots,k_n)\,
d\psp_m(q;k_1,\ldots,k_m)\,,
\label{ps:phspfact}
\eeq
which simply follows from inserting the identity
\beq
1\equiv 
\int \frac{d^dq}{(2\pi)^d}
(2\pi)^d \delta^{(d)}\left(q-\sum_{i=1}^m k_i\right)
\eeq
in eq.~(\ref{ps:phspdef}). In the context of higher-order QCD computations, 
the phase space decomposition of eq.~(\ref{ps:phspfact}) matches the
asymptotic behaviour of the matrix elements in the singular limits,
which are expressed as a product of a kernel that collects all kinematic
singularities, times a reduced non-divergent matrix element. In an NNLO
computation, up to three parton momenta will enter the kernel. Thus,
we shall use eq.~(\ref{ps:phspdef}) with $m=2,3$.

We start from the two-body phase space $d\psp_2(q;k_1,k_2)$. In order to 
give a frame-independent phase-space parametrization, we express the
angular measure in eq.~(\ref{ps:Emeas}) in terms of invariants.
In order to do this, two reference light-like four-vectors need be introduced,
which define a system of axes; when azimuthal integration is trivial,
one of these vectors becomes irrelevant. After defining the reference
vectors (which we denote by $n$ and $t$, the latter relevant only to
azimuthal integration), the angles which enter 
eqs.~(\ref{ps:dOmegaone})--(\ref{ps:dOmegak}) are written in terms of 
four-momenta through Gram determinants (see sect.~2 of ref.~\cite{Byckling}):
\beq
\Delta_n(p_1,\dots,p_n)\equiv \left | 
\begin{array}{cccc}
p_1^2 & p_1\mydot p_2 & ... & p_1\mydot p_n\\
p_2\mydot p_1 & p_2^2 & ... & p_2\mydot p_n\\
...   & ... & ... & ...\\
p_n\mydot p_1 & p_n\mydot p_2 & ...& p_n^2\\
\end{array}
\right |\,.
\eeq
We also define the variables
\beqn
z_i&=&\frac{k_i\mydot n}{q\mydot n}\,,\;\;\;\;\;\;i=1,2\,,
\label{ps:zidef}
\\
s_{12}&=&2k_1\mydot k_2\,.
\label{ps:sijdef}
\eeqn
After a somewhat lenghty algebra we arrive at
\beqn
d\psp_2(q;k_1,k_2)&=&
\frac{\Omega^{(d-3)}}{8(2\pi)^{d-2}}\,
\frac{\Delta_2(n,q)^{\frac{d-3}{2}}\Delta_4(n,k_1,k_2,t)^{\frac{d-5}{2}}}
{\left(\Delta_3(n,k_1,k_2)\Delta_3(n,q,t)\right)^{\frac{d-4}{2}}}
\left(-\frac{\Delta_3(n,k_1,k_2)}{\Delta_2(n,q)}\right)^{\frac{d-4}{2}}
\nonumber \\*&\times&
\delta\left(1-z_1-z_2\right)
\delta\left(q^2-s_{12}-k_1^2-k_2^2\right)
dz_1 dz_2 ds_{12} ds_{1t}\,,
\label{ps:twobdazi}
\eeqn
where the second (third) fraction on the r.h.s. is due to the azimuthal
(polar) part of the angular measure. By carrying out the azimuthal 
integration, we get
\beqn
d\psp_2(q;k_1,k_2)&=&
\frac{\Omega^{(d-2)}}{4(2\pi)^{d-2}}\,
\left(-\frac{\Delta_3(n,k_1,k_2)}{\Delta_2(n,q)}\right)^{\frac{d-4}{2}}
\nonumber \\*&\times&
\delta\left(1-z_1-z_2\right)
\delta\left(q^2-s_{12}-k_1^2-k_2^2\right)
dz_1 dz_2 ds_{12}\,.
\label{ps:twobd}
\eeqn
Using eqs.~(\ref{ps:zidef}) and~(\ref{ps:sijdef}) we obtain
\beq
-\frac{\Delta_3(n,k_1,k_2)}{\Delta_2(n,q)}=
z_1 z_2 q^2 - z_1 k_2^2 - z_2 k_1^2\,.
\label{ps:twobdgram}
\eeq
We now proceed analogously for the three-body phase space
$d\psp_3(q;k_1,k_2,k_3)$. In this case, we shall not need to consider
massive partons, and we thus set $k_i^2=0$ from the beginning.
Furthermore, one of the two azimuthal integrations will always be 
carried out trivially in our computations; thus, it is not necessary
to introduce the auxiliary vector $t$, since its role can be played
by one of the four-momenta $k_i$ (which was not possible in the
two-body case). We get
\beqn
&&d\psp_3(q;k_1,k_2,k_3)=
\frac{\Omega^{(d-2)}\Omega^{(d-3)}}{32(2\pi)^{2d-3}}
\left(\frac{\Delta_4(n,k_1,k_2,k_3)}{\Delta_2(n,q)}\right)^{\frac{d-5}{2}}
\nonumber \\&&\phantom{d\psp_{1\to 3}}\times
\delta\left(1-z_1-z_2-z_3\right) 
\delta\left(q^2-s_{12}-s_{13}-s_{23}\right)
dz_1 dz_2 dz_3 ds_{12} ds_{13} ds_{23}\,,\phantom{aaaa}
\label{ps:threebd}
\eeqn
where the variables $z_i$ and $s_{ij}$ have been defined similarly
to eqs.~(\ref{ps:zidef}) and~(\ref{ps:sijdef}). Upon explicit 
computation of the Gram determinants, we obtain
\beqn
\frac{\Delta_4(n,k_1,k_2,k_3)}{\Delta_2(n,q)}&=&
\frac{1}{4}\Big(2 z_1 z_2 s_{13} s_{23}+
2 z_1 z_3 s_{12} s_{23}+2 z_2 z_3 s_{12} s_{13}
\nonumber\\&&\phantom{\frac{1}{4}}
-z_1^2 s_{23}^2-z_2^2 s_{13}^2-z_3^2 s_{12}^2\Big)\,.
\eeqn
Equation~(\ref{ps:threebd}) is fully symmetric in $k_i$, and is therefore
suited to treat the most general $1\to 3$ parton branching. There are
cases in which such a symmetry is absent due to physical reasons (for
example, a strongly-ordered double-collinear limit, i.e. $C\oplus C$);
in order to treat these, there are more convenient parametrizations
than that of eq.~(\ref{ps:threebd}). We use the factorization
formulae eq.~(\ref{ps:phspfact}) with $n=3$, $m=2$, and
\beq
\begin{array}{rcl}
q & \longrightarrow & p+k_3 \\
  &  & \!\bentarrow {k_1+k_2}\,,
\end{array}
\label{ps:momtbstrong}
\eeq
and we get
\beq
d\psp_3(q;k_1,k_2,k_3)=
\frac{dp^2}{2\pi}\,d\psp_2(q;k_3,p)\,d\psp_2(p;k_1,k_2)\,,
\label{ps:threebdso}
\eeq
where for the phase spaces on the r.h.s. we can use 
eqs.~(\ref{ps:twobdazi}) and~(\ref{ps:twobd}). In particular, upon
introducing the variables
\beqn
&&z_i=\frac{k_i\mydot n}{q\mydot n}\,,\;\;\;\;\;
z_{12}=\frac{p\mydot n}{q\mydot n}\,,\;\;\;\;\;
s_{3(12)}=2k_3\mydot p\,,
\label{ps:zdso}
\\
&&\zeta_1=\frac{k_1\mydot n}{p\mydot n}\,,\;\;\;\;\;
\zeta_2=\frac{k_2\mydot n}{p\mydot n}\,,\;\;\;\;\;
\label{ps:zidso}
\eeqn
and using eq.~(\ref{ps:twobd}) and eq.~(\ref{ps:twobdazi}) we obtain
\beqn
d\psp_2(q;k_3,p)&=&
\frac{\Omega^{(d-2)}}{4(2\pi)^{d-2}}\,
\left(-\frac{\Delta_3(n,k_3,p)}{\Delta_2(n,q)}\right)^{\frac{d-4}{2}}
\nonumber \\*&\times&
\delta\left(1-z_3-z_{12}\right)
\delta\left(q^2-s_{3(12)}-p^2\right)
dz_3 dz_{12} ds_{3(12)}\,,
\label{ps:tdsoinn}
\eeqn
and
\beqn
d\psp_2(p;k_1,k_2)&=&
\frac{\Omega^{(d-3)}}{8(2\pi)^{d-2}}\,
\frac{\Delta_2(n,p)^{\frac{d-3}{2}}\Delta_4(n,k_1,k_2,k_3)^{\frac{d-5}{2}}}
{\left(\Delta_3(n,k_1,k_2)\Delta_3(n,k_1+k_2,k_3)\right)^{\frac{d-4}{2}}}
\left(-\frac{\Delta_3(n,k_1,k_2)}{\Delta_2(n,p)}\right)^{\frac{d-4}{2}}
\nonumber \\*&\times&
\delta\left(1-\zeta_1-\zeta_2\right)
\delta\left(p^2-s_{12}\right)
d\zeta_1 d\zeta_2 ds_{12} ds_{13}\,.
\label{ps:tdsoout}
\eeqn
Upon explicit computation of the Gram determinants, we get
\beqn
-\frac{\Delta_3(n,k_3,p)}{\Delta_2(n,q)}&=&
z_3 z_{12} q^2 - z_3 p^2,
\label{ps:gramsa}
\\
-\frac{\Delta_3(n,k_1,k_2)}{\Delta_2(n,p)}&=&
\zeta_1 \zeta_2 p^2,
\label{ps:gramsb}
\eeqn
which are analogous to eq.~(\ref{ps:twobdgram}). From eqs.~(\ref{ps:zdso})
and~(\ref{ps:zidso}) we also get
\beq
z_{12}=z_1+z_2\,,\;\;\;\;\;\;
\zeta_j=\frac{z_j}{1-z_3}\,.
\eeq
We can introduce the azimuthal variable $x$ through
\beqn
\label{ps:exkin}
s_{13}&=&(s_{123}-s_{12})(1-\zeta_2)+\f{\sqrt{s_{12}}}{1-z_3}\xi\,,
\label{ps:sot}
\\
s_{23}&=&(s_{123}-s_{12})\zeta_2-\f{\sqrt{s_{12}}}{1-z_3}\xi\,,
\label{ps:stt}
\eeqn
where\footnote{Note that the variable $\xi$ is directly related to the 
combination $z_2 s_{13}-z_1s_{23}$ appearing in $t_{12,3}$. We have 
$z_2 s_{13}-z_1s_{23}=\sqrt{s_{12}}\xi$.}
\beq
\xi=2\sqrt{z_3 \zeta_2(1-\zeta_2)(s_{123}(1-z_3)-s_{12})}\, 
x+z_3(2\zeta_2-1)\sqrt{s_{12}}\,.
\label{ps:exxi}
\eeq
Using these definitions we rewrite the three-body phase space of
eq.~(\ref{ps:threebdso}) as follows
\beqn
d\psp_3&=&\f{\Omega^{(d-2)}\Omega^{(d-3)}}{16(2\pi)^{2d-3}}
\left(s_{123}\right)^{\f{d-4}{2}}\left(s_{12}\right)^{\f{d-4}{2}}
\left(1-\f{s_{12}}{s_{123}(1-z_3)}\right)^{\f{d-4}{2}}
\left(z_3(1-z_3)\right)^{\f{d-4}{2}}
\nonumber \\&\times&
\left(\zeta_2\,(1-\zeta_2)\right)^{\f{d-4}{2}}\,
(1-x^2)^{\f{d-5}{2}}\, ds_{12}\, dz_3\, d\zeta_2\, dx\,.
\label{ps:pstbso}
\eeqn
We stress that eqs.~(\ref{ps:threebd}) and~(\ref{ps:pstbso}) are
identical, i.e. no approximation has been made in the latter. In
fact, eqs.~(\ref{ps:pstbso}) could be
obtained from eq.~(\ref{ps:threebd}) by simply changing integration
variables.

\end{document}